\documentclass[fleqn]{article}
\usepackage[english]{babel}
\usepackage{a4wide}
\usepackage{latexsym}
\usepackage{times}
\usepackage{theorem}
\usepackage{url}
\usepackage[final]{graphics}
\usepackage{amsmath,amssymb}
\usepackage{amsfonts}
\usepackage{array}
\usepackage{calc}
\usepackage{xspace}
\usepackage{color}
\usepackage{epsfig}
\usepackage{subfigure}
\usepackage{wrapfig}
\usepackage{float}
\usepackage{stmaryrd}
\usepackage{color}
\usepackage{mathtools}
\usepackage{tikz}
\usetikzlibrary{calc,arrows}
\usetikzlibrary{decorations.pathreplacing}
\usepackage{hyperref}
\usepackage{booktabs}

\newcounter{theoremcnt}[section]
\renewcommand{\thetheoremcnt}{\thesection.\arabic{theoremcnt}}

{\begin{trivlist}\refstepcounter{theoremcnt}
\item[]\bf Definition~\thetheoremcnt.~\rm}{\end{trivlist}}

\newenvironment{definition-arg}[1]%
{\begin{trivlist}\refstepcounter{theoremcnt}
\item[]\bf Definition~\thetheoremcnt ~(#1).~\rm}{\end{trivlist}}

{\begin{trivlist}\refstepcounter{theoremcnt}
\item[]\bf Example~\thetheoremcnt.~\rm}{\end{trivlist}}

{\begin{trivlist}\refstepcounter{theoremcnt}
\item[]\bf Lemma~\thetheoremcnt.~\rm}{\end{trivlist}}
{\begin{trivlist}\item[]{\bf Proof.}}{\qed\end{trivlist}}

{\begin{trivlist}\refstepcounter{theoremcnt}
\item[]\bf Theorem~\thetheoremcnt.~\rm}{\end{trivlist}}

{\begin{trivlist}
\item[]{\bf Acknowledgements.} }{\end{trivlist}}

\newcolumntype{L}{>{$}l<{$}}%stopzone%stopzone%stopzone
\newcolumntype{C}{>{$}c<{$}}%stopzone%stopzone%stopzone
\newcolumntype{R}{>{$}r<{$}}%stopzone%stopzone%stopzone

{\begin{trivlist}
\item\begin{tabular}{@{}>{\bf}p{2.3em}L@{\ }L@{\ }L@{\ }L@{\ }L@{\ }L@{\ }L@{\ }L}}%
{\end{tabular}\end{trivlist}}

\newcommand{\qed}{~\hfill$\Box$}
\newcommand{\Real}{\mathbb{R}}
\newcommand{\Nat}{\mathbb{N}}

\newcommand{\Int}{\mathbb{Z}}
\newcommand{\Bool}{\mathbb{B}}
\newcommand{\dataeq}{\approx}
\newcommand{\true}{\mathit{true}}
\newcommand{\false}{\mathit{false}}

\title{Formal Modelling and Analysis of Slot Machines}
\author{Jan Friso Groote, Sander van Heesch and Matthias Volk\\
{\footnotesize\sl Departement of Mathematics and Computer Science,
Eindhoven University of Technology}\\
{\footnotesize\sl P.O.~Box 513, 5600 MB Eindhoven, The Netherlands}\\
{\footnotesize \sl Email: \tt \{J.F.Groote, M.Volk\}@tue.nl, S.J.A.G.v.Heesch@student.tue.nl}}
\date{}

\begin{document}
\maketitle
\begin{abstract}
\noindent
Slot machines can have fairly complex behaviour. 
Determining the \textit{RTP} (\textit{return to player}) can be involved, especially when a player has an influence on the course of 
the game.
In this paper we model the behaviour of slot machines using probabilistic process specifications 
where the intervention of players is modelled using non-determinism. 
The RTP is formulated as a quantitative modal formula which can be evaluated fully automatically on the behavioural specifications of these slot machines.
We apply the method on an actual slot machine provided by the company Err\`{e}l Industries B.V. 
The most useful contribution of this paper is that we show how to
describe the behaviour of slot machines both concisely and unequivocally.
Using quantitative modal logics there is an extra bonus, as 
we can quite easily provide valuable insights by
a.o.\ computing the exact RTP and obtaining the optimal player strategies. 
\end{abstract}

\section{Introduction}
\label{sec:intro}
Gambling machines such as slot machines combine random behaviour, stemming from spinning of reels, with player input, when allowing to hold specific reels.
A major metric for such machines is the payout, also called the `Return To Player' (RTP).
The RTP is the average part of the inserted money which is returned to the player in each round.
A lower bound on the RTP is commonly required by law or regulation and differs per country.
For example, in the Netherlands slot machines must have an average payout of at least 60\% and do not allow players to loose more than 40 euros per hour (\cite[Articles 12f and 12g]{Wetboek}; see also \cite{NedWet}).
It is of course also undesirable for the owners of such machines when the average payout exceeds what is inserted, i.e., the RTP should be less than 100\%.
Determining the RTP is therefore crucial when designing slot machines.
However, as these machines can be quite complex, it can be difficult to assess the RTP with traditional mathematical methods.
Monte Carlo simulation can be quite effective in such situation, but it remains a probabilistic simulation technique and can require a large number of simulation runs.

In this paper, we model the slot machines using probabilistic process specifications and formulate the property of interest, such as the RTP, using quantitative modal formulas~\cite{GM14}.
By using process descriptions quite complex behaviour can be concisely described.
These models can be analysed fully automatically using model checking of 
modal formulas and this yields precise answers.
For example, we can check whether the expected payout lies within the required bounds.
In contrast to traditional mathematical methods, model checking is performed fully automatically on the model specification and requires no user intervention.
In contrast to simulation-based techniques, model checking exhaustively explores all possible behaviour and therefore also considers rare events which only occur with marginal probability.

\paragraph{Process specifications}
Process algebras are very natural formalisms to describe and study the behaviour of complex systems.
These formalisms find their origin in the work of Milner~\cite{Milner80}.
Systems can perform actions, such as displaying information or receiving an indication that a button has been pressed.
Abstractly, actions are denoted as $a$ and $b$, and more informatively as $\mathit{display}(1)$ or $\mathit{press}$. 
A process algebra expression specifies all possible sequences in which such actions can occur, which we call the behaviour.
In this paper we use the process algebraic specification language mCRL2 \cite{GM14}.

Modal formulas allow to formulate a wide range of properties for the behaviour of the modelled system.
A typical example is that `whenever action $a$ happens, action $b$ will inevitably take place'.
We use the modal mu-calculus with data, which is very expressive, in particular more expressive than other modal logics \cite{CranenGroote11}. 
Using model checking techniques, we check whether the given modal formula is satisfied on the given process specification.
In other words, we verify whether the modelled system ensures the desired behaviour.
Tools and theories that support process descriptions and allow to verify properties in the modal mu-calculus are CADP~\cite{CaesarAldebaran} and mCRL2~\cite{mCRL2}. 

\paragraph{Probabilistic process specifications}
The process algebraic descriptions have been extended to incorporate probabilities. 
Using this it can for instance be expressed that `after an action $a$ with probability of 25\% an action $b$ will happen, and with probability 75\% both actions $b$ and $c$ are possible'.
This allows to specify and investigate behaviour of systems that rely on probabilities. 

Quantitative modal logics yield numerical values instead of the booleans true and false, when evaluated on behaviour.
Using these logics various properties can be establish such as the probability that a particular action $a$ will happen, or the expected number of times that action $a$ will happen.
There are various tools that support probabilistic process behaviour and allow to verify properties such as Prism~\cite{Prism}, Storm~\cite{Storm} and Modest~\cite{Modest}.
We employ the recent quantitative extension of the modal mu-calculus as outlined in~\cite{GrooteWillemse2023} which has been implemented in the mCRL2 toolset~\cite{mCRL2}.

\paragraph{Modeling slot machines}
We start out by modelling a number of simple slot machines to illustrate the approach.
Subsequently, we apply the techniques to the TopSpinner slot machine produced by Err\`{e}l Industries B.V.\ \cite{JVH}.
This machine was produced in 2001 and is not in commercial use anymore.
We model the TopSpinner in mCRL2 and analyse its behaviour via quantitative model checking.

It turns out that modelling slot machines is very well possible, once one masters the technique of process algebraic specification and quantitative model checking.
Two of the three non-trivial games of the TopSpinner neatly produce a return to player of 94\% and our final model of the third game has a payout of 83\%. 

However, using the available documentation it was hardly possible to 
figure out how this slot machine operated precisely. 
It required a number of clarification
rounds, and even the use of a simulator of the actual slot machine, to be able to provide
precise formal descriptions. 
We initially started with this project to investigate the effectiveness
of quantitative modal logic in the setting of slot machines, and that 
turned out to work fine.
But we came to the conclusion that the formal modelling of the behaviour of slot machines per se is a much larger contribution. 
Formal modelling is perfectly suited to denote what slot machines are 
supposed to do, which besides as a basis for analysis, is an excellent
way to communicate this behaviour with for instance gambling authorities.

\paragraph{Related work}
There is remarkable little literature on slot machines, given their
relative popularity and nice probabilistic nature. 
Very simple slot machines have been used as illustration, see e.g., \cite{DBLP:journals/tcs/BradfieldS92,Mao2012}.
The mathematics of slot machines has been investigated in~\cite{Barboianu2013}.
Analysis of slot machines via Monte Carlo simulation was considered in~\cite{BarrDurbach2008}.
One goal of slot machine analysis is optimizing the RTP such that it falls within the desired bounds.
Several approaches exist for this optimization problem, based on genetic and evolutionary algorithms~\cite{Balabanov2015B,Balabanov2015,Keremedchiev2017} or using Variable Neighborhood Search~\cite{Kamanas2021}.

\paragraph{Outline}
We introduce the process specification mCRL2 and quantitative model logic in Section~\ref{sec:specification}.
We model and analyse simple slot machines in Section~\ref{sec:simple_slot_machines} before investigating the different games of the Top Spinner in Section~\ref{sec:top_spinner}.
We conclude in Section~\ref{sec:conclusion}.

\paragraph{Acknowledgements.} We are thankful to Eric van de Pas from Eurocoin gaming for his kind patience
in answering all our questions about slot machines.

\section{The specification formalisms}
\label{sec:specification}
We describe the behaviour of systems in mCRL2, a (probabilistic) process algebraic behavioural specification language \cite{GM14}. 
Quantitative modal logic is used to express probabilistic properties of this behaviour, such as the
expected payout of a slot machine \cite{GrooteWillemse2023}. In the next two sections we explain the necessary
ingredients of both formalisms. 
\subsection{Probabilistic process expressions}
Probabilistic process specifications consist of two parts, namely data types and behaviour.
\paragraph{Data types}
All normal data types can be used, such as booleans~($\Bool$), natural numbers~($\Nat$), integers~($\Int$), real numbers~($\Real$), and lists~($\mathit{List}(D)$ where $D$ is an arbitrary type).
Lists are typically denoted within square brackets, such as $[1,2,3]$. The $i$-th element of a list $L$ is
denoted as $L.i$, where the first element of the list is $L.0$. 

Enumerated types are typically declared using the 
$\textbf{struct}$ keyword, as follows:
\[\begin{array}{l@{\hspace{2mm}}l}
\mathbf{sort}&\mathit{Symbol}=\textbf{struct }\mathit{star}\mid \mathit{grapes}\mid\mathit{orange};
\end{array}\]
where $\mathit{Symbol}$ is now the data type containing the three elements $\mathit{star}$, 
$\mathit{grapes}$ and $\mathit{orange}$.

Auxiliary functions and constants are declared using the $\textbf{map}$ keyword, followed by defining equations, employing the
keyword $\textbf{eqn}$ and the keyword $\textbf{var}$ to declare the variables used in the equations. An example to determine
whether three symbols are all equal is the following where $\approx$ represents equality on data elements. 
\[\begin{array}{l@{\hspace{2mm}}l}
\textbf{map}&\mathit{all\_equal}:\mathit{Symbol}\times\mathit{Symbol}\times\mathit{Symbol}\rightarrow\Bool;\\
\textbf{var}&s_1,s_2,s_3:\mathit{Symbol};\\
\textbf{eqn}&\mathit{all\_equal}(s_1,s_2,s_3)=~~s_1\approx s_2 \wedge s_2\approx s_3;
\end{array}\]
The equations are evaluated as term rewriting rules from left to right. They can be preceded by a condition, in which case
the rewrite rule is only applied if the condition rewrites to true. Within equations we often use the ternary operator $\mathit{if}(b,x,y)$
to represent the value $x$ if the boolean $b$ is true and $y$ if $b$ is equal to false. 

The data types are far more versatile, also allowing for instance function types, sets, bags and recursive types, but as they are
not required in the rest of this paper, we do not explain those.

\paragraph{Behaviour}
Behaviour is described in the process algebraic style. The essential element in process algebra is the action, typically denoted
in an abstract way by $a$, $b$, $c$. They represent some atomically occurring event. If the event has some meaning, then
more informative action names can be used such as $\mathit{win}$ or $\mathit{lose}$. Actions can have parameters, which are then
denoted after the action within brackets. A typically example is $\mathit{payout}(n)$ which typically could 
indicate the atomic activity of paying out $n$ credits, or $\mathit{hold}(b_1,b_2,b_3)$ with three booleans that indicate which of 
the three hold buttons are pressed. Actions are declared using the keyword $\textbf{act}$.
\[\begin{array}{l@{\hspace{2mm}}l}
\textbf{act}&\mathit{win}, \mathit{lose};\\
&\mathit{hold}:\Bool\times\Bool\times\Bool;
\end{array}\]

The operator `${\cdot}$' is used to sequentially compose two behaviours. E.g., $\mathit{win}{\cdot}\mathit{lose}$ represents that first the
action $\mathit{win}$ takes place, followed by the action $\mathit{lose}$. 
The operator $+$ represents non-deterministic choice between two behaviours. For instance $a{\cdot} b+ c{\cdot}d$ means that it is
either possible to do $a{\cdot}b$ or do $c{\cdot}d$ where the first action, $a$ or $c$, that is done determines the choice. 
There is one process, called deadlock, written as $\delta$, which cannot perform any action. It typically satisfies the
equation $p+\delta=\delta+p=p$ for any process $p$. 

The operator `$+$' can also be written using the sum operator $\sum$. Consider a process $p(b)$, which is a process $p$ 
depending on a boolean $b$. Suppose we want to express that either the behaviour $p(\true)$ or $p(\false)$ can
be done. This is expressed by $p(\true)+p(\false)$, but using the the sum operator it can be written as $\sum_{b:\Bool}p(b)$.
The sum operator can be used with any data type, and with one or multiple variables. 

Using the conditional process operator $b\rightarrow p\diamond q$, it is possible to let data influence the course of a process. 
If $b$ is true the behaviour of $p$ is executed, and if $b$ is false the behaviour of $q$ takes place. Note that the conditionals
on data and processes have different notation, but they have essentially the same effect. When omitting the else-part, i.e., $b\rightarrow p$
this process executes $p$ when $b$ is true, and is equal to deadlock otherwise. 

Using process equations, iterative or recursive behaviour can be denoted. A simple example is the 
behaviour of the process $\mathit{Lucky}$ which can perform the action $\mathit{win}$ followed by the
behaviour of $\mathit{Lucky}$. So, $\mathit{Lucky}$ can do an infinite sequence of actions $\mathit{win}$. 
The process starts with the behaviour indicated by the keyword $\textbf{init}$.
\[
\begin{array}{@{}ll}
\begin{array}{l@{\hspace{2mm}}l}
\textbf{proc}&\mathit{Lucky}=\mathit{win}{\cdot}\mathit{Lucky};\\
\textbf{init}&\mathit{Lucky};
\end{array}\hspace*{3.5cm}&
\raisebox{-0.8cm}[0cm][0cm]{
\begin{tikzpicture}{c}
%Triangles and I's and V's.
\draw [thick,->] (-0.3,0.0) -- (-0.1,0.0);
\draw [thick] (0,0) circle(0.1);
\draw [thick,->] (0.07,0.07) .. controls (1,1) and (1,-1) .. (0.07,-0.07);
\draw (1.2,0) node {$\mathit{win}$};
%\draw [thick] (0,0.2) -- (-1.7,3) -- (1.7,3) -- (0,0.2);
\end{tikzpicture}}
\end{array}
\]
The keyword $\textbf{proc}$ is used to indicate the process equation. At the left side of the equals sign a single process variable occurs of
which the behaviour occurs at the right hand side. 

Such behaviour is interpreted as a labelled transition system, also called a state machine. It has a set of states and transitions
labelled with actions indicating how the behaviour goes from state to state. The state machine for the process $\mathit{Lucky}$ is drawn at the right of $\mathit{Lucky}$.
It consists of one state and a transition labelled with the action $\mathit{win}$ going from and to this state indicating that
the action $\mathit{win}$ can be done infinitely often in sequence.

Process equations can also have zero or more parameters. As an illustration we write the behaviour of a counter that can count up
and down, but not count below zero. The counter initially starts with the value~$0$. 
\[
\begin{array}{@{}ll}
\begin{array}{l@{\hspace{2mm}}l}
\textbf{act}&\mathit{up},\mathit{down};\\
\textbf{proc}&\mathit{Count}(n:\Int)=\\
             &\hspace{1cm}\mathit{up}{\cdot}\mathit{Count}(n{+}1)+\\
             &\hspace{1cm}(n>0)\rightarrow\mathit{down}{\cdot}\mathit{Count}(n{-}1);\\
\textbf{init}&\mathit{Count}(0);
\end{array}&\hspace*{1cm}
\raisebox{-0.7cm}[0cm][0cm]{
\begin{tikzpicture}
%Triangles and I's and V's.
\draw [thick,->] (-0.3,0.0) -- (-0.1,0.0);
\draw [thick] (0,0) circle(0.1);
\draw [thick] (1,0) circle(0.1);
\draw [thick] (2,0) circle(0.1);
\draw [thick, gray] (3,0) circle(0.1);
\draw [thick, lightgray] (4,0) circle(0.1);
\draw [lightgray] (4.5,0) node{$\cdots$};
\draw [thick,->] (0.07,0.07) .. controls (0.5,0.3) .. (0.93,0.07);
\draw (0.5,0.4) node {$\mathit{up}$};
\draw [thick,->] (1.07,0.07) .. controls (1.5,0.3) .. (1.93,0.07);
\draw (1.5,0.4) node {$\mathit{up}$};
\draw [thick,->,gray] (2.07,0.07) .. controls (2.5,0.3) .. (2.93,0.07);
\draw [gray] (2.5,0.4) node {$\mathit{up}$};
\draw [thick,->, lightgray] (3.07,0.07) .. controls (3.5,0.3) .. (3.93,0.07);
\draw [lightgray] (3.5,0.4) node {$\mathit{up}$};
\draw [thick,<-] (0.07,-0.07) .. controls (0.5,-0.3) .. (0.93,-0.07);
\draw (0.5,-0.4) node {$\mathit{down}$};
\draw [thick,<-] (1.07,-0.07) .. controls (1.5,-0.3) .. (1.93,-0.07);
\draw (1.5,-0.4) node {$\mathit{down}$};
\draw [thick,<-,gray] (2.07,-0.07) .. controls (2.5,-0.3) .. (2.93,-0.07);
\draw [gray] (2.5,-0.4) node {$\mathit{down}$};
\draw [thick,<-,lightgray] (3.07,-0.07) .. controls (3.5,-0.3) .. (3.93,-0.07);
\draw [lightgray] (3.5,-0.4) node {$\mathit{down}$};
\end{tikzpicture}}
\end{array}
\]
Note that the operator $+$ is used here both as the addition on integers and as the non-deterministic choice between processes. 
Also observe that if $n=0$ the action $\mathit{down}$ cannot be done, but the action $\mathit{up}$ is still possible. 
The labelled transition system belonging to this process is infinite and is partly drawn at the right of the process specification.

Probabilistic behaviour is indicated by $\textbf{dist }d{:}D[\mathit{dist}].p$ which expresses that the behaviour $p$ is executed
with a value for the variable $d$ which is chosen according to the distribution $\mathit{dist}$. The distribution $\mathit{dist}$ must
sum up to one over all elements of the distribution. As a concrete example we
express that one of the symbols $\mathit{star}$, $\mathit{grapes}$ and $\mathit{orange}$ is displayed, each with probability $\frac{1}{3}$
using the following process specification. In this case we only allow to gamble once, and we indicate that by putting $\delta$ after the 
display action. 
\[
\begin{array}{@{}ll}
\begin{array}{l@{\hspace{2mm}}l}
\textbf{act}&\mathit{display}:\mathit{Symbol};\\
\textbf{proc}&\mathit{Gamble}=\textbf{dist }s{:}\mathit{Symbol}[\frac{1}{3}].\mathit{display}(s){\cdot}\delta;\\
\textbf{init}&\mathit{Gamble};             
\end{array}&\hspace*{1cm}
\raisebox{-1.1cm}[0cm][1cm]{
\begin{tikzpicture}
%Triangles and I's and V's.
\draw [thick,->] (1,2.3) -- (1,2.1);
\draw [thick] (1,2) circle(0.1);
\draw [thick] (0,1) circle(0.1);
\draw [thick] (1,1) circle(0.1);
\draw [thick] (2,1) circle(0.1);
\draw [thick] (0,0) circle(0.1);
\draw [thick] (1,0) circle(0.1);
\draw [thick] (2,0) circle(0.1);
%Probabilistic transitions.
\draw [thick,->] (0.93,1.93) -- (0.07,1.07);
\draw (0.3,1.7) node {$\frac{1}{3}$};
\draw [thick,->] (1,1.90) -- (1,1.1);
\draw (1.2,1.4) node {$\frac{1}{3}$};
\draw [thick,->] (1.07,1.93) -- (1.93,1.07);
\draw (1.7,1.7) node {$\frac{1}{3}$};
\draw [thick] (0.8,1.8) .. controls (1,1.7) .. (1.2,1.8);
%Action transitions
\draw [thick,->] (0,0.93) -- (0,0.1);
\draw (-0.8,0.6) node{\scriptsize$\mathit{display}(\mathit{star})$};
\draw [thick,->] (1,0.93) -- (1,0.1);
\draw (1,0.4) node{\scriptsize$\mathit{display}(\mathit{grapes})$};
\draw [thick,->] (2,0.93) -- (2,0.1);
\draw (3.0,0.6) node{\scriptsize$\mathit{display}(\mathit{orange})$};
\end{tikzpicture}}
\end{array}
\]

The transition system drawn at the right is a probabilistic transition system. In this case, with probability $\frac{1}{3}$
one of the states at the mid level are selected, from which actions can be done. This two layer behaviour, first
probabilities and then actions, can be repeated to indicate repeated probabilistic choices. When there is only
one single probabilistic transition with probability $1$ in a state, then it is generally omitted and the action transitions
are drawn immediately in that state. 

Besides probabilistic behaviour it is also possible to model non-deterministic behaviour, which is behaviour that
is fully unpredictable. For instance, when a slot machine has hold buttons to fix a particular symbol on the screen
it is unpredictable which buttons a player will press. Note that this pressing behaviour is not random, as the
player can or cannot have a particular strategy. Using the `$+$' or $\sum$ operators such non-determinism can be expressed.
We extend the previous example with a hold button where $\mathit{hold}(\true)$ means that the hold button is 
pressed, and $\mathit{hold}(\false)$ means the symbol is not fixed and can be randomly set again. Note that the equation
for $\mathit{Gamble}$ models the game where the symbol needs to be randomly chosen, and $\mathit{Gamble}(s)$ models
the game where the symbol is set to $s$. 
\[
\begin{array}{@{}ll}
\begin{array}{l@{\hspace{2mm}}l}
\textbf{act}&\mathit{display}:\mathit{Symbol};\\
            &\mathit{hold}:\Bool;\\
\textbf{proc}&\mathit{Gamble}=\textbf{dist }s{:}\mathit{Symbol}[\frac{1}{3}].\mathit{Gamble}(s);\\
             &\mathit{Gamble}(s{:}\mathit{Symbol})=\\
             &\hspace{0.5cm}\mathit{display}(s){\cdot}\\
             &\hspace*{1cm}(\mathit{hold}(\true){\cdot}\mathit{Gamble}(s)+{}\\
             &\hspace*{1.15cm}\mathit{hold}(\false){\cdot}\mathit{Gamble});\\
\textbf{init}&\mathit{Gamble};             
\end{array}&\hspace*{-0.5cm}
\raisebox{-2cm}[0cm][1cm]{
\begin{tikzpicture}
%Nodes.
\draw [thick,->] (1,2.3) -- (1,2.1);
\draw [thick] (1,2) circle(0.1);
\draw [thick] (0,1) circle(0.1);
\draw [thick] (1,1) circle(0.1);
\draw [thick] (2,1) circle(0.1);
\draw [thick] (0,0) circle(0.1);
\draw [thick] (1,-1.5) circle(0.1);
\draw [thick] (2,0) circle(0.1);
%Probabilistic transitions.
\draw [thick,->] (0.93,1.93) -- (0.07,1.07);
\draw (0.3,1.7) node {$\frac{1}{3}$};
\draw [thick,->] (1,1.90) -- (1,1.1);
\draw (1.2,1.4) node {$\frac{1}{3}$};
\draw [thick,->] (1.07,1.93) -- (1.93,1.07);
\draw (1.7,1.7) node {$\frac{1}{3}$};
\draw [thick] (0.8,1.8) .. controls (1,1.7) .. (1.2,1.8);
%Display transitions
\draw [thick,->] (0,0.93) -- (0,0.1);
\draw (-0.8,0.6) node{\scriptsize$\mathit{display}(\mathit{star})$};
\draw [thick,->] (1,0.93) -- (1,-1.4);
\draw (0,-0.7) node{\scriptsize$\mathit{display}(\mathit{grapes})$};
\draw [thick,->] (2,0.93) -- (2,0.1);
\draw (3.0,0.6) node{\scriptsize$\mathit{display}(\mathit{orange})$};
%Hold transitions
\draw [thick,->] (-0.1,0) .. controls (-2.2,0.4) and (-2.2,0.8) .. (-0.1,1);
\draw (-1,1.1) node{\scriptsize$\mathit{hold}(\true)$};
\draw [thick,->] (-0.1,0) .. controls (-2.6,-0.2) and (-2.6,2.0) .. (0.9,2);
\draw (-1.2,2.0) node{\scriptsize$\mathit{hold}(\false)$};
%second
\draw [thick,->] (1.07,-1.43) .. controls (1.4,0) .. (1.07,0.93);
\draw (2.0,-0.4) node{\scriptsize$\mathit{hold}(\true)$};
\draw [thick,->] (0.9,-1.5) .. controls (-3,-1.8) and (-3.4,2.2) .. (0.9,2);
\draw (-1,-1.6) node{\scriptsize$\mathit{hold}(\false)$};
%third
\draw [thick,->] (2.1,0) .. controls (4.7,0.1) and (4.7,1.2) .. (2.1,1);
\draw (3,1.2) node{\scriptsize$\mathit{hold}(\true)$};
\draw [thick,->] (2.1,0) .. controls (5.1,-0.2) and (5.1,2.0) .. (1.1,2);
\draw (3.2,2.0) node{\scriptsize$\mathit{hold}(\false)$};
\end{tikzpicture}}
\end{array}
\]

The language mCRL2 offers more ways of describing behaviour, in particular to describe parallel communicating 
components and time. We refer to \cite{GM14} for the details.

\subsection{Quantitative modal logic}
Modal logics are developed to state and evaluate properties about behaviour. There are various modal logics, 
but we use Hennessy-Milner logic \cite{HennessyMilner} extended with fixed point operators and data \cite{GM14}, called the modal
$\mu$-calculus, as this is the most expressive modal logic available \cite{CranenGroote11}. 
In this logic each formula is evaluated on a state of a 
transition system  and interpreted as true or false. 

In \cite{Gawlitza1, Gawlitza2, GrooteWillemse2023} this logic is extended and interpreted on $\Real_{\{-\infty,\infty\}}$, i.e., each formula
yields a real number including $\pm \infty$.
Such a formula can be a real number, $\true$, interpreted as $\infty$, $\false$, interpreted as $-\infty$. There are
operators $+$, $\wedge$, $\vee$ where $\wedge$ is interpreted as the minimum and $\vee$ represents maximum. It is 
also possible to multiply a formula $\phi$ with a positive constant as in $c*\phi$. A valid formula is $((\frac{3}{9}*2)+1)\wedge\true$,
which on any transition system is interpreted as $\frac{5}{3}$. 

Using the diamond modality $\langle a\rangle\phi$ and the box modality $[a]\phi$ the logic is connected to a probabilistic
transition system. These modalities stem from \cite{HennessyMilner}. The evaluation of a formula in a state with probabilistic
transitions is the value obtained by evaluating the formula in the target states multiplied with the probability of the transition.

The evaluation of the diamond modality $\langle a\rangle\phi$ in a state with outgoing action transitions is the maximum of
the evaluation of $\phi$ in all states reachable via an action $a$. If there are no outgoing $a$-transitions, the interpretation
is $-\infty$.
For instance, the evaluation of $0\vee(\langle \mathit{display}(\mathit{grapes})\rangle 1)$ in the transition system of the process $\mathit{Gamble}$ is $\frac{1}{3}$, because the probability to perform action $\mathit{display}(\mathit{grapes})$ is $\frac{1}{3}$. 
Using the explicit values $0$ and $1$ it evaluates to $1$ if the action $\mathit{display}(\mathit{grapes})$ is possible in 
a state, and to $0$ if not. Using the probabilities, this value $1$ contributes $\frac{1}{3}$ to the result. 
Note that the formula $\langle \mathit{display}(\mathit{grapes})\rangle 1$ would evaluate to $-\infty$ because the formula
evaluates to $-\infty$ in any state without outgoing transition $\mathit{display}(\mathit{grapes})$. Such states occur at
the left and the right. These $-\infty$-ies dominate the outcome which is also $-\infty$. 

The box modality $[a]\phi$ does the same as the diamond modality, except that it takes the minimum of the evaluation
of $\phi$ over all $a$-transitions that it can do. If there are no $a$-transitions the box modality evaluates to infinity.
Note that in states with exactly one outgoing transition labelled with action $a$, the diamond modality $\langle a \rangle \phi$ and $[a]\phi$ evaluate to exactly the same value. 

Instead of an action, it is also possible to put a regular expression between the modality operator, such as
$\langle \true\rangle\phi$ and $\langle \true^\star\rangle\phi$. The $\true$ represents any action and $\true^\star$
stands for any finite sequence of actions. For instance $\langle \true^\star \rangle(\langle\mathit{win}\rangle1\vee 0)$
is the probability that the action $\mathit{win}$ is reachable. 

Using the infimum and supremum operators, $\inf d{:}D.\phi$ and $\sup d{:}D.\phi$, respectively, 
it is possible to minimise or maximise the value of a formula over the elements
of a data domain. For instance, $\sup s{:}\mathit{Symbol}.\langle \mathit{display}(s)\rangle\phi$ provides 
the maximal value of evaluating this formula with $\mathit{star}$, $\mathit{grapes}$ and $\mathit{orange}$. 

Using the minimal fixed point operator $\mu$ and maximal fixed point operator $\nu$ values for iterative behaviour can
be obtained. Concretely, consider a minimal fixed point formula $\mu X.\phi$ where $X$ can occur in $\phi$.
If this formula is evaluated in a state, it yields the smallest value $v$ such that $v$ equals
$\phi$ with $X$ in $\phi$ set to $v$. When the maximimum fixed point formula is used, it yields the largest such value $v$.
For instance, the formula $\nu X.\langle \mathit{win}\rangle X$ is equal to $\infty$ if it is evaluated on a transition 
system that can perform an infinite sequence of actions $\mathit{win}$ as in the process $\mathit{Lucky}$. 
Otherwise, it evaluates to $-\infty$. 

In fixed point operators, parameters can be used. For instance, the formula $\mu X(n{:}\Nat{=}0).(n\vee\langle \mathit{win}\rangle X(n+1))$
counts the maximal number of consecutive actions $\mathit{win}$. It provides the smallest value $n$, being at least $0$,
and at least one more for every consecutive action $\mathit{win}$ that can be done. If an infinite number of 
actions $\mathit{win}$ can be done, this formula yields $\infty$.

\section{Simple slot machines}
\label{sec:simple_slot_machines}
We show how to make models of slot machines of increasing complexity. The behaviour of the slot machine is systematically
described in the behavioural language of mCRL2. Assessment of winning probabilities and `return to player' are obtained
via modal formulas. 

\subsection{A one column slot machine}
We start with a simple slot machine showing only one symbol selected out of a star, 
grapes and an orange with equal
probability. 
When a star is shown the player wins one credit.
In the other two cases he loses and it costs one credit.
Directly after winning or losing, the player can play again, and this can be repeated indefinitely. 
\[
\begin{array}{@{}ll}
\begin{array}{ll}
\textbf{sort }&\mathit{Symbol}=\textbf{struct }\mathit{star}\mid\mathit{grapes}\mid\mathit{orange};\\
\textbf{act }&\mathit{win},\mathit{lose};\\
             &\mathit{display}:\mathit{Symbol};\\
\textbf{proc }&\mathit{Play} = \textbf{dist }s{:}\mathit{Symbol}[1/3].\mathit{display}(s){\cdot}\\
              &\hspace*{2cm}((s\approx \mathit{star})\rightarrow\mathit{win}{\cdot}\mathit{Play}+{}\\
              &\hspace*{2cm}\phantom{(}(s\approx \mathit{grapes}\vee s\approx\mathit{orange})\rightarrow\mathit{lose}{\cdot}\mathit{Play});\phantom{)}\\
\textbf{init }&\mathit{Play};
\end{array}&\hspace*{1cm}
\raisebox{-0.7cm}[0cm][1cm]{
\begin{tikzpicture}
%Triangles and I's and V's.
\draw [thick] (0,0) -- (0,1.62) -- (1,1.62) -- (1,0) -- (0,0);
\draw [thick] (-0.1,-0.1) -- (-0.1,1.72) -- (1.1,1.72) -- (1.1,-0.1) -- (-0.1,-0.1);
% Draw a star.
\filldraw [thick, yellow] (0.88,0.92) -- (0.12,0.92) -- (0.74,0.48) -- (0.5,1.2) -- (0.26,0.48) -- (0.88,0.92);
\draw  (0.88,0.92) -- (0.12,0.92) -- (0.74,0.48) -- (0.5,1.2) -- (0.26,0.48) -- (0.88,0.92);
\end{tikzpicture}}
\end{array}\]
The probabilistic transition system for the one column slot machine is given below. 
\begin{center}
\begin{tikzpicture}
%Triangles and I's and V's.
\draw [thick,->] (1,2.3) -- (1,2.1);
\draw [thick] (1,2) circle(0.1);
\draw [thick] (0,1) circle(0.1);
\draw [thick] (1,1) circle(0.1);
\draw [thick] (2,1) circle(0.1);
\draw [thick] (0,0) circle(0.1);
\draw [thick] (1,0) circle(0.1);
\draw [thick] (2,0) circle(0.1);
%Probabilistic transitions.
\draw [thick,->] (0.93,1.93) -- (0.07,1.07);
\draw (0.3,1.7) node {$\frac{1}{3}$};
\draw [thick,->] (1,1.90) -- (1,1.1);
\draw (1.2,1.4) node {$\frac{1}{3}$};
\draw [thick,->] (1.07,1.93) -- (1.93,1.07);
\draw (1.7,1.7) node {$\frac{1}{3}$};
\draw [thick] (0.8,1.8) .. controls (1,1.7) .. (1.2,1.8);
%Display transitions
\draw [thick,->] (0,0.93) -- (0,0.1);
\draw (-0.8,0.6) node{\scriptsize$\mathit{display}(\mathit{star})$};
\draw [thick,->] (1,0.93) -- (1,0.1);
\draw (1,0.4) node{\scriptsize$\mathit{display}(\mathit{grapes})$};
\draw [thick,->] (2,0.93) -- (2,0.1);
\draw (3.0,0.6) node{\scriptsize$\mathit{display}(\mathit{orange})$};
%Win/lose transitions
\draw [thick,->] (-0.1,0) .. controls (-2.6,0.0) and (-2,1.9) .. (0.9,2);
\draw (-1.4,1.6) node{\scriptsize$\mathit{win}$};
\draw [thick,->] (1.07,-0.07) .. controls (6.6,-0.7) and (4,2.3) .. (1.1,2);
\draw (3.5,1.8) node{\scriptsize$\mathit{lose}$};
\draw [thick,->] (2.1,0) .. controls (5.6,0.0) and (4,1.7) .. (1.1,2);
\draw (3,1.3) node{\scriptsize$\mathit{lose}$};
\end{tikzpicture}
\end{center}
The probability to win a single game is given by the modal formula
\[\sup s{:}\mathit{Symbol}.\langle \mathit{display}(s)\rangle(\langle \mathit{win}\rangle 1 \vee 0).\]
The answer is $\frac{1}{3}$. 

We can replace $\mathit{display}(s)$ by $\true$ 
\begin{equation}
\label{expected_once}
\langle \true\rangle(\langle \mathit{win}\rangle 1 \vee 0),
\end{equation}
which also yields $\frac{1}{3}$. It says that after an arbitrary action, $1$ is returned if $\mathit{win}$ is 
possible, and otherwise $0$. These values are weighted with the probabilities in the distribution.

Replacing $\mathit{display}(s)$ by $\true^\star$ changes the meaning. The formula now becomes
\[\langle \true^\star\rangle(\langle \mathit{win}\rangle 1 \vee 0).\]
It gives the probability of reaching $\mathit{win}$ after an arbitrary sequence of actions. This essentially provides
the probability that on the one column slot machine star will ultimately appear. This probability is of course $1$ as is confirmed
by evaluating the formula. 

The expected number of times one has to play to encounter a star is given by the formula
\begin{equation}
\label{expected_nr_of_steps}
\mu X.\sup s{:}\mathit{Symbol}.\langle\mathit{display}(s)\rangle(s\approx\mathit{star} \wedge 1) \vee 
((s\not\approx \mathit{star} \wedge \langle\true\rangle (X+1)) \vee 0).
\end{equation}
This formula expresses that the expected number of rounds to encounter a star is equal to $1$ if the star is encountered
directly, and it is $1$ plus the expected number of rounds if no star is observed in a round. We search for the
minimum value of the fixed point $X$ that is at least zero. If zero is not added, the minimal solution is $-\infty$. 
The formula evaluates to $3$ meaning that one encounters a star after $3$ rounds on average. 

The one column slot machine behaves exactly the same for every round. But suppose we would be interested in the average long run reward per round, this can be formulated by:
\begin{equation}
\label{longrun}
\begin{array}{@{}l}
   \mu X(\mathit{gain}{:}\Int{=}0, \mathit{rounds}{:}\Nat{=}0).\\
   \hspace*{1cm}\mathit{rounds}\approx\mathit{max\_rounds} \wedge \mathit{gain}/\mathit{rounds} \vee{}\\
   \hspace*{1cm}\mathit{rounds}<\mathit{max\_rounds} \wedge{}\\
   \hspace*{2cm}( \langle\true\rangle\langle\mathit{win}\rangle X(\mathit{gain}{+}1,\mathit{rounds}{+}1) \vee
                 \langle\true\rangle\langle\mathit{lose}\rangle X(\mathit{gain}{-}1, \mathit{rounds}{+}1)
               ).
\end{array}
\end{equation}
We calculate this gain explicitly for $\mathit{max\_rounds}$ of rounds. In the variable $\mathit{gain}$ the cumulative
gain for all rounds is recalled. If the number of rounds $\mathit{rounds}$ becomes equal to the maximal number
of rounds, the average gain $\mathit{gain}/\mathit{rounds}$ is returned. Otherwise, if the game is won, indicated
by the modalities $\langle \true\rangle\langle\mathit{win}\rangle\ldots$, the expected value with the gain incremented
by $1$ is delivered, and if the game is lost, shown by $\langle \true\rangle\langle\mathit{lose}\rangle\ldots$,
the expected value with the gain decremented by $1$ is the result. Note that the use of variables in fixed
point operators allow for far more complex results to be derived on behaviour. As expected, evaluating this formula
on the behaviour of the one column slot machine yields an average gain of $-\frac{1}{3}$.
In Section \ref{sec:reelgame} a variant of this formula
is used where only one parameter in the fixed point variable is required. 
\subsection{A three column slot machine}
\label{threecolumn}
A three column slot machine has three windows where stars, grapes and oranges can 
appear each with equal probability. A game is won if in all three windows the same symbol appears.
Its behaviour is described using the following mCRL2 specification. Note the use of if-then-else, denoted
by $\ldots\rightarrow\ldots\diamond\ldots$, which describes winning if the symbols are equal, and losing, otherwise.
\[
\begin{array}{@{}ll}
\begin{array}{ll}
\textbf{sort }&\mathit{Symbol}=\textbf{struct }\mathit{star}\mid\mathit{grapes}\mid\mathit{orange};\\
\textbf{act }&\mathit{win},\mathit{lose};\\
             &\mathit{display}:\mathit{Symbol}\times\mathit{Symbol}\times\mathit{Symbol};\\
\textbf{proc }&\mathit{Play} = \textbf{dist }s_1,s_2,s_3{:}\mathit{Symbol}[1/27].\mathit{display}(s_1,s_2,s_3){\cdot}\\
              &\hspace*{2cm}((s_1\approx s_2\wedge s_2\approx s_3)\rightarrow\mathit{win}\diamond\mathit{lose}){\cdot}\mathit{Play};\\
\textbf{init }&\mathit{Play};
\end{array}&\hspace*{0.2cm}
\raisebox{-0.7cm}[0cm][1cm]{
\begin{tikzpicture}
%Triangles and I's and V's.
\draw [thick] (0,0) -- (0,1.62) -- (1,1.62) -- (1,0) -- (0,0);
\draw [thick] (-0.1,-0.1) -- (-0.1,1.72) -- (1.1,1.72) -- (1.1,-0.1) -- (-0.1,-0.1);
% Draw a star.
\filldraw [thick, yellow] (0.88,0.92) -- (0.12,0.92) -- (0.74,0.48) -- (0.5,1.2) -- (0.26,0.48) -- (0.88,0.92);
\draw  (0.88,0.92) -- (0.12,0.92) -- (0.74,0.48) -- (0.5,1.2) -- (0.26,0.48) -- (0.88,0.92);
\end{tikzpicture}
\begin{tikzpicture}
%Triangles and I's and V's.
\draw [thick] (0,0) -- (0,1.62) -- (1,1.62) -- (1,0) -- (0,0);
\draw [thick] (-0.1,-0.1) -- (-0.1,1.72) -- (1.1,1.72) -- (1.1,-0.1) -- (-0.1,-0.1);
% Draw an orange.
\filldraw [thick, orange] (0.5,0.81) circle(0.4);
\draw (0.5,0.81) circle(0.4);
\draw (0.75,1.0) circle(0.05);
\end{tikzpicture}
\begin{tikzpicture}
%Triangles and I's and V's.
\draw [thick] (0,0) -- (0,1.62) -- (1,1.62) -- (1,0) -- (0,0);
\draw [thick] (-0.1,-0.1) -- (-0.1,1.72) -- (1.1,1.72) -- (1.1,-0.1) -- (-0.1,-0.1);
% Draw an orange.
\filldraw [thick, orange] (0.5,0.81) circle(0.4);
\draw (0.5,0.81) circle(0.4);
\draw (0.75,1.0) circle(0.05);
\end{tikzpicture}}
\end{array}\]
We can essentially analyse this game with the same formulas as in the previous section. The probability
to win the game once, given by formula (\ref{expected_once}) is $\frac{1}{9}$. 
The formula (\ref{expected_nr_of_steps}) to indicate the expected number of steps to win must slightly be
adapted, as winning now means seeing three similar symbols. It becomes:
\begin{equation}
\label{wininfirstrounddisplay}
\begin{array}{@{}l}
   \mu X.\sup s_1,s_2,s_3{:}\mathit{Symbol}.\langle\mathit{display}(s_1,s_2,s_3)\rangle\\
\hspace*{3cm} (s_1\approx s_2 \wedge s_2\approx s_3 \wedge 1 \vee
                  ((s_1\not\approx s_2 \vee s_2\not\approx s_3) \wedge \langle \true\rangle(X{+}1)) \vee 0)
\end{array}
\end{equation}
and yields the answer $9$.

The average expected loss per game is given by the formula (\ref{longrun}) and it yields $-\frac{7}{9}$.

\subsection{A three column slot machine with hold buttons}
Slot machines sometimes have hold buttons. After a game in which nothing was won, the player
can press the hold buttons under each symbol, to fix these symbols for the next game. For a three column slot
machine this is beneficial for the player, because when there are two equal symbols that are fixed, the probability
of winning the next game increases to $\frac{1}{3}$. 

The behaviour of such as slot machine becomes more complicated. Not only the winning probability can differ 
for each game, but also there is non-deterministic behaviour as the player may or may not press the
hold buttons. 

The slot machine with hold buttons can be described as follows. 
\[
\begin{array}{@{}ll}
\begin{array}{ll}
\textbf{sort }&\mathit{Symbol}=\textbf{struct }\mathit{star}\mid\mathit{grapes}\mid\mathit{orange};\\
\textbf{map }&\mathit{distribution}: \Bool\times \mathit{Symbol}\times \mathit{Symbol}\rightarrow \Real;\\
\textbf{var }&b: \Bool;\\
    &r, s: \mathit{Symbol};\\
\textbf{eqn }&\mathit{distribution}(b, r, s) = \mathit{if}(b, \mathit{if}(r\dataeq s, 1, 0), \frac{1}{3});\\
\textbf{act }&\mathit{win},\mathit{lose};\\
             & \mathit{hold}:\Bool\times\Bool\times\Bool;\\
             &\mathit{display}:\mathit{Symbol}\times\mathit{Symbol}\times\mathit{Symbol};\\
\textbf{proc }&\mathit{Play}(\mathit{hold}_1, \mathit{hold}_2, \mathit{hold}_3:\Bool, r_1, r_2, r_3: \mathit{Symbol}) =\\
       &\hspace*{1cm}\textbf{dist }s_1,s_2,s_3:\mathit{Symbol}[\mathit{distribution}(\mathit{hold}_1, r_1, s_1)*\\
       &\hspace*{4.4cm}\mathit{distribution}(\mathit{hold}_2, r_2, s_2)*\\
       &\hspace*{4.4cm}\mathit{distribution}(\mathit{hold}_3, r_3, s_3)].\\
       &\hspace*{2cm}\mathit{display}(s_1,s_2,s_3){\cdot}\\
       &\hspace*{2cm}(s_1\dataeq s_2 \wedge s_2\dataeq s_3)\\
       &\hspace*{2.3cm}\rightarrow\mathit{win}{\cdot}\mathit{Play}(\false, \false, \false,s_1,s_2,s_3)\\
       &\hspace*{2.3cm}~\diamond~~\mathit{lose}{\cdot}
       \sum_{b_1, b_2, b_3: \Bool}
       \mathit{hold}(b_1,b_2,b_3){\cdot}\mathit{Play}(b_1, b_2, b_3, s_1, s_2, s_3);\\

\textbf{init }&\mathit{Play}(\false, \false, \false, \mathit{star}, \mathit{star}, \mathit{star});
\end{array}&\hspace*{-2cm}
\raisebox{-0cm}[0cm][1cm]{
\begin{tikzpicture}
%Box
\draw [thick] (0,0) -- (0,1.62) -- (1,1.62) -- (1,0) -- (0,0);
\draw [thick] (-0.1,-0.1) -- (-0.1,1.72) -- (1.1,1.72) -- (1.1,-0.1) -- (-0.1,-0.1);
\draw [thick] (0.5,-0.6) circle(0.3) node{\footnotesize hold};
% Draw an orange.
\filldraw [thick, orange] (0.5,0.81) circle(0.4);
\draw (0.5,0.81) circle(0.4);
\draw (0.75,1.0) circle(0.05);
\end{tikzpicture}
\begin{tikzpicture}
%Box
\draw [thick] (0,0) -- (0,1.62) -- (1,1.62) -- (1,0) -- (0,0);
\draw [thick] (-0.1,-0.1) -- (-0.1,1.72) -- (1.1,1.72) -- (1.1,-0.1) -- (-0.1,-0.1);
\draw [thick] (0.5,-0.6) circle(0.3) node{\footnotesize hold};
% Draw a star.
\filldraw [thick, yellow] (0.88,0.92) -- (0.12,0.92) -- (0.74,0.48) -- (0.5,1.2) -- (0.26,0.48) -- (0.88,0.92);
\draw  (0.88,0.92) -- (0.12,0.92) -- (0.74,0.48) -- (0.5,1.2) -- (0.26,0.48) -- (0.88,0.92);
\end{tikzpicture}
\begin{tikzpicture}
%Box
\draw [thick] (0,0) -- (0,1.62) -- (1,1.62) -- (1,0) -- (0,0);
\draw [thick] (-0.1,-0.1) -- (-0.1,1.72) -- (1.1,1.72) -- (1.1,-0.1) -- (-0.1,-0.1);
\draw [thick] (0.5,-0.6) circle(0.3) node{\footnotesize hold};
% Draw grapes.
\filldraw [thick, blue] (0.5,0.4) circle(0.1);
\draw (0.5,0.4) circle(0.1);
\filldraw [thick, blue] (0.4,0.5) circle(0.1);
\draw (0.4,0.5) circle(0.1);
\filldraw [thick, blue] (0.6,0.5) circle(0.1);
\draw (0.6,0.5) circle(0.1);
\filldraw [thick, blue] (0.35,0.7) circle(0.1);
\draw (0.35,0.7) circle(0.1);
\filldraw [thick, blue] (0.65,0.69) circle(0.1);
\draw (0.65,0.69) circle(0.1);
\filldraw [thick, blue] (0.5,0.7) circle(0.1);
\draw (0.5,0.7) circle(0.1);
\filldraw [thick, blue] (0.3,0.9) circle(0.1);
\draw (0.3,0.9) circle(0.1);
\filldraw [thick, blue] (0.73,0.9) circle(0.1);
\draw (0.73,0.9) circle(0.1);
\filldraw [thick, blue] (0.3,1.1) circle(0.1);
\draw (0.3,1.1) circle(0.1);
\filldraw [thick, blue] (0.4,1.15) circle(0.1);
\draw (0.4,1.15) circle(0.1);
\filldraw [thick, blue] (0.7,1.1) circle(0.1);
\draw (0.7,1.1) circle(0.1);
%%%%%%%%%%%%%% upper row 3
\filldraw [thick, blue] (0.6,1.14) circle(0.1);
\draw (0.6,1.14) circle(0.1);
%%%%%%%%%%%%%% second upper row 2 and 3
\filldraw [thick, blue] (0.45,0.93) circle(0.1);
\draw (0.45,0.93) circle(0.1);
\filldraw [thick, blue] (0.55,0.92) circle(0.1);
\draw (0.55,0.92) circle(0.1);

\end{tikzpicture}
\begin{tikzpicture}
\end{tikzpicture}}

\end{array}\]
The behaviour is given by the process $\mathit{Play}$ with $6$ parameters. The first three parameters 
$\mathit{hold}_1$, $\mathit{hold}_2$ and $\mathit{hold}_3$ indicate which of the hold buttons are pressed.
The last three parameters represent the symbols shown on the screen. Initially, the hold buttons cannot
be used, as indicated by $\false$,$\false$,$\false$. In this case the initial symbols on the screen are not
relevant, and they are arbitrarily set to $\mathit{star}$,  $\mathit{star}$, $\mathit{star}$. Note that after
the action $\mathit{win}$, the hold buttons are also disabled. 

When playing, the symbols are set using the mapping $\mathit{distribution}(b,r,s)$ where $b$ is a boolean and
$r$ and $s$ are symbols. This mapping determines the probability to display symbol $s$ in a single column. If
boolean $b$ is true, the hold button for this column is pressed and $s$ is equal to the pre-set symbol $r$ with
probability $1$. Otherwise, $s$ is equal to some concrete symbol with probability $\frac{1}{3}$. Using the action 
$\mathit{display}(s_1,s_2,s_3)$ it is indicated which symbols are shown on the screen.

Clearly, if all symbols are the same, the player wins one credit, indicated by the action $\mathit{win}$. 
Otherwise, the player loses a credit shown by the action $\mathit{lose}$, after which the player can indicate which
hold buttons should be pressed. Subsequently, the player can continue to play.

Using the formula (\ref{wininfirstrounddisplay}) it is determined what the probability is to win the
game in the first round, which is $\frac{1}{9}$ leading to an expected loss of $-\frac{7}{9}$ credits in the 
first round. But on the one hand if the hold buttons are used well, the winning probability can be increased and the
average loss can be restricted. On the other hand, if the hold buttons are used very badly, the loss is increased.
To get insight in the effect of the hold buttons, we investigate three situation, namely first where the hold buttons
are used perfectly, second where the hold buttons are used in the worst possible manner, and third where the hold
buttons are used randomly.
We summarize the results in Table~\ref{tab:three_reel_game}.

\begin{table}[b]
\centering
\caption{Strategy comparison for three column slot machine with hold buttons.}
\label{tab:three_reel_game}

\begin{tabular}{lrr}
  \toprule
  Strategy & Winning probability & Expected loss\\
  \midrule
	Best   & 0.2591 & -0.4818\\
	Worst  & 0.0001 & -0.9998\\
	Random & 0.0886 & -0.8227\\
  \bottomrule
\end{tabular}
\end{table}%

\paragraph{Best strategy}
The following formula corresponds to the maximal winning probabilities per round where the player uses
the hold buttons in the best possible way. Again we use the constant $\mathit{max\_rounds}$ to set the maximum
number of rounds over which the average is calculated. 
\begin{equation}
\label{maxhold}
\begin{array}{@{}l}
\mu X(\mathit{wins}{:}\Nat{=}0, \mathit{rounds}{:}\Nat{=}0).\\
\hspace*{1cm}\mathit{rounds}{\approx}\mathit{max\_rounds} \wedge \mathit{wins}/\mathit{rounds} \vee{}\\
\hspace*{1cm}\mathit{rounds}{<}\mathit{max\_rounds} \wedge{}
( \langle\true\rangle\langle \mathit{win}\rangle X(\mathit{wins}{+}1,\mathit{rounds}{+}1) \vee{}\\
\hspace*{4.3cm}  \langle\true\rangle\langle\mathit{lose}\rangle\langle\true\rangle X(\mathit{wins}, \mathit{rounds}{+}1)
               ).
\end{array}
\end{equation}
The sequence $\langle\true\rangle\langle \mathit{win}\rangle$ matches an action $\mathit{display}$ followed by 
an action $\mathit{win}$. Similarly, the sequence 
$\langle\true\rangle\langle\mathit{lose}\rangle\langle\true\rangle$ matches
a $\mathit{display}$, followed by $\mathit{lose}$, followed by an action $\mathit{hold}$. This action hold
is non-deterministic, and by using the diamond modality $\langle \true\rangle\ldots$ we take the maximal winning
probability over the various possible settings of the hold buttons. Setting $\mathit{max\_rounds}$ to $1000$
yields a winning probability of $26\%$ per round.
This corresponds to an expected loss of $-0.48$ credits per round.

\paragraph{Worst strategy}
By replacing the last modality in $\langle\true\rangle\langle\mathit{lose}\rangle\langle\true\rangle$ in formula
(\ref{maxhold}) by $[\true]$ the minimal expected winning probability is obtained which corresponds to the situation
where a player presses the hold buttons in the worst possible way. Evaluating this formula over $1000$ rounds,
yields a winning probability of $0.0125\%$.
This corresponds to the situation where after the player lost a game, the player continuous to press all hold buttons. This causes a certain loss of the game in each round, explaining the success rate of close to $0\%$ and an expected loss of $-1$ per round. 

\paragraph{Random strategy}
We may also be interested in the situation where a player plays `naturally', in the sense that he does not play
optimally nor perfectly badly. The question is how to model `natural' behaviour. One way is to assume that the player
randomly presses the hold buttons. The average winning probability for this situation is expressed in the following
formula. 
\begin{equation}
\nonumber
\begin{array}{@{}l}
\mu X(\mathit{wins}{:}\Nat{=}0, \mathit{rounds}{:}\Nat{=}0).\\
\hspace*{1cm}\mathit{rounds}{\approx}\mathit{max\_rounds} \wedge \mathit{wins}/\mathit{rounds} \vee{}\\
\hspace*{1cm}\mathit{rounds}{<}\mathit{max\_rounds} \wedge
               ( \langle\true\rangle\langle\mathit{win}\rangle X(\mathit{wins}{+}1,\mathit{rounds}{+}1) \vee{}\\
\hspace*{4.3cm}\langle \true\rangle\langle\mathit{lose}\rangle \sum_{b_1,b_2,b_3{:}\Bool}.
        \frac{1}{8}*\langle \mathit{hold}(b_1,b_2,b_3)\rangle X(\mathit{wins}, \mathit{rounds}{+}1)
               );
\end{array}
\end{equation}
In the last line it is expressed using the sum operator that each pattern of pressing the hold buttons should
be considered as appearing with probability $\frac{1}{8}$. This formula then evaluates to an average winning 
probability of $9\%$ and an average loss per round of $-0.82$ credits.
Note that by taking a more elaborate formula for pressing the hold buttons, more delicate behaviour can be 
modelled, for instance by only allowing to press the buttons when equal symbols are shown.

\section{Top Spinner from Err\`{e}l Industries B.V.}
\label{sec:top_spinner}
\begin{figure}
\begin{center}
\includegraphics[width=1.2\textwidth, angle=-90]{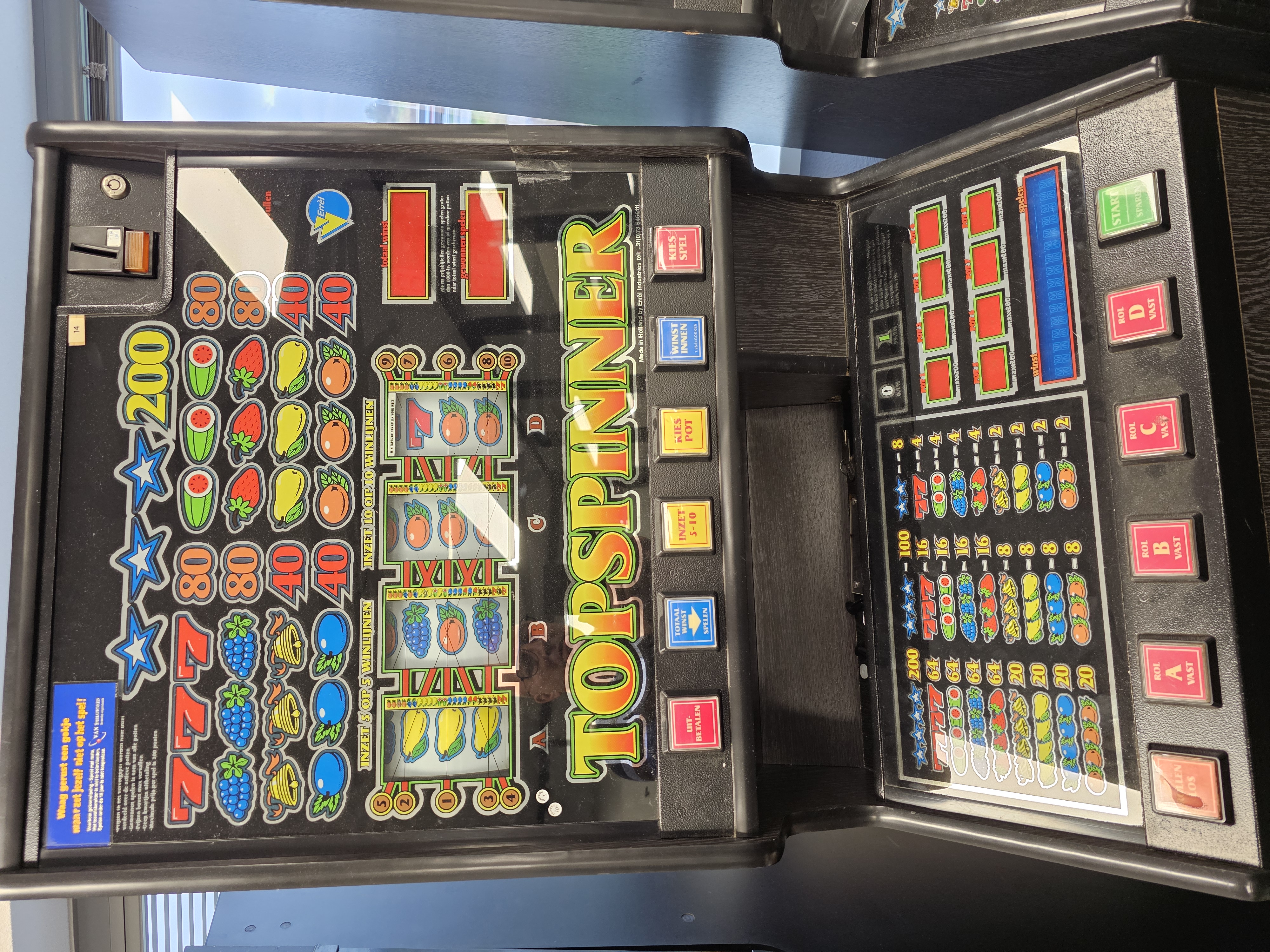}
\end{center}
\caption{The Top Spinner slot machine by Err\`{e}l Industries B.V.}
\label{TopSpinner}
\end{figure}
In Figure \ref{TopSpinner} a slot machine called the Top Spinner produced by Err\`{e}l Industries is depicted\footnote{A 
similar, online playable version is available at \cite{OnlineGame}}.
It has a playing window of three rows and four columns, four corresponding hold buttons and contains four games. Initially, the 
reel game is played of which the winning combinations are depicted at the bottom of the machine. The payout of the reel
game is moved to the pots. Alternatively, 
the zero-one-game is used to transfer credits to one of the eight pots. This game is very
simple and corresponds to the zero and one on the bottom part of the cabinet in the right upper corner.
If a pot contains
more than 5 or 10 credits, the five-play-line or ten-play-line game can also be played, respectively, although it is obligatory to play
the zero-one game once before a play-line game can be played. 
If all pots contain more than 5 credits the reel and zero-one games are not available.
The winning scheme of the two play-line games are depicted at the top of the machine. 
We describe all four games separately below, and analyse the three most complex ones. 
\subsection{The zero-one game}
The zero-one game is used to transfer money to the pot and it must sometimes be played before the
play-line games can be done. With probability of 99.9\% a credit is successfully transferred,
and otherwise it is lost. This game is so simple that we do not analyse it further. 
\subsection{The five-play-lines game}
\begin{figure}
\[\begin{array}{ll@{\hspace{-1.0cm}}r}
\textbf{sort}&\mathit{Symbol} = \textbf{struct } \mathit{orange} \mid \mathit{grapes} \mid \mathit{pear} \mid \mathit{melon} \mid 
\mathit{blueberry} \mid \mathit{strawberry} \mid \mathit{bell} \mid \mathit{seven} \mid \mathit{star};\\
&     \mathit{Reel} = \mathit{List}(\mathit{Symbol});\\
\textbf{map}
     &r_1, r_2, r_3, r_4: \mathit{Reel};\\
&     \mathit{get}_1,\mathit{get}_2,\mathit{get}_3, \mathit{get}_4:\Nat \rightarrow \mathit{Symbol};\\
&\mathit{distribution}:\Nat\rightarrow\Real;\\
\textbf{eqn}&  r_1 = [\mathit{orange},\mathit{orange},\mathit{star},\mathit{orange},\mathit{grapes},\mathit{grapes},
\mathit{pear},\mathit{pear},\\
&\hspace*{1cm}\mathit{pear},\mathit{pear},\mathit{melon},\mathit{melon},\mathit{blueberry},\mathit{blueberry},
\mathit{blueberry},\mathit{blueberry},\\
&\hspace*{1cm}\mathit{strawberry},\mathit{strawberry},\mathit{bell},\mathit{bell},\mathit{bell},
\mathit{bell},\mathit{seven},\mathit{seven}];\\
&     r_2 = [\mathit{orange},\mathit{orange},\mathit{orange},\mathit{grapes},\mathit{orange},\mathit{grapes},\mathit{pear},\mathit{star},\mathit{pear},\\
&\hspace*{1cm}\mathit{melon},\mathit{pear},\mathit{melon},\mathit{blueberry},\mathit{blueberry},\mathit{blueberry},\mathit{strawberry},\mathit{blueberry},\\
&\hspace*{1cm}\mathit{strawberry},\mathit{bell},\mathit{bell},\mathit{bell},\mathit{seven},\mathit{bell},\mathit{seven}];\\
&     r_3 = [\mathit{orange},\mathit{orange},\mathit{orange},\mathit{orange},\mathit{grapes},\mathit{grapes},\mathit{pear},\mathit{pear},\\
&\hspace{1cm}\mathit{pear},\mathit{pear},\mathit{melon},\mathit{melon},\mathit{blueberry},\mathit{blueberry},\mathit{star},\mathit{blueberry},\\
&\hspace*{1cm}\mathit{strawberry},\mathit{strawberry},\mathit{bell},\mathit{bell},\mathit{bell},\mathit{bell},\mathit{seven},\mathit{seven}];\\
&r_4 = [\mathit{orange},\mathit{orange},\mathit{orange},\mathit{orange},\mathit{grapes},\mathit{grapes},\mathit{pear},\mathit{pear},\\ &\hspace{1cm}\mathit{pear},\mathit{pear}, \mathit{melon},\mathit{melon},\mathit{blueberry},\mathit{blueberry},\mathit{blueberry},\mathit{blueberry},\\
&\hspace{1cm} \mathit{strawberry},\mathit{strawberry},\mathit{bell},\mathit{bell},\mathit{star},\mathit{bell},\mathit{seven}, \mathit{seven}];\\
\textbf{var} & i: \Nat;\\
\textbf{eqn}&     \mathit{distribution}(i) = \mathit{if}(i{<}24, 1/24, 0);\\
&     \mathit{get}_1(i)=r_1.(i\,\mathrm{mod}\,24);\\
&     \mathit{get}_2(i)=r_2.(i\,\mathrm{mod}\,24);\\
&     \mathit{get}_3(i)=r_3.(i\,\mathrm{mod}\,24);\\
&     \mathit{get}_4(i)=r_4.(i\,\mathrm{mod}\,24);\\

\end{array}\]
\caption{The mCRL2 data types for the five-play-lines and ten-play-lines games}
\label{datatypesfiveandten}
\end{figure}

The five-play-line game requires five credits to play. It corresponds to the winning lines indicated with 1 to 5
at the upper part of the cabinet. Each 
winning line has length 3 and considers columns A--C. If one of the combinations at the top occurs at a winning line, 
then the corresponding amount is payed out. For instance, the combination star-star-star yields 200 credits, 
and three bells yield 40. If multiple winning lines show winning combinations, the payout is added up.
The stars have a double role. A star in column one is also an orange, a star in column two doubles as a pear, 
a star in the third column is also a blueberry, and a star in the last column is also a bell. So, the combination
star-orange-orange is also good for 40 credits, as are pear-star-pear and blueberry-blueberry-star. 
The combination bell-bell-star is not relevant for the five-play-line game, as 
the five winning lines only use the left three columns. 

The game does not use the hold buttons, and in that sense is similar to the three column slot machine
from Section \ref{threecolumn}, albeit more complex. The expected reward for each round is exactly the same, and therefore 
it suffices to analyse a single round of the game. 

The behavioural model in mCRL2 is provided in Figures \ref{datatypesfiveandten} and \ref{fiveplaylines}. 
In Figure \ref{datatypesfiveandten} the basic data types are provided that are both used in the five-play-lines game
and the ten-play-lines games. The sort $\mathit{Symbol}$ contains the nine symbols that play a role. Each reel is modelled
as a list of length 24. The reels are $r_1$, $r_2$, $r_3$ and $r_4$, counted from left to right. 
So, $r_1$ is the left most reel, containing an orange, an orange, a star, etc. 

The functions $\mathit{get}_j(i)$ get the $i$-th symbol from reel $j$. As each reel has length 24, a symbol at position $i$
is put at the lowest line of the display of the slot machine with probability $\frac{1}{24}$, provided $i<24$. This
is defined in the mapping $\mathit{distribution}$. 

\begin{figure}[t]
\[\begin{array}{ll@{\hspace{-1.0cm}}r}
\textbf{map}&      \mathit{price}: \mathit{Symbol}\times\mathit{Symbol}\times\mathit{Symbol} \rightarrow \Nat;\\
  &   \mathit{reward}: \Nat\times\Nat\times\Nat \rightarrow \Nat;\\
\textbf{var} & i_1, i_2, i_3: \Nat;\\
\textbf{eqn}&  
     \mathit{price}(s_1,s_2,s_3)=\mathit{if}(s_1\approx \mathit{star} \wedge s_2\approx\mathit{star} \wedge s_3\approx \mathit{star}, 200,\\
&\hspace*{2.8cm}  \mathit{if}(s_1\approx s_2 \wedge s_2\approx s_3 \wedge s_1 \in \{ \mathit{seven}, \mathit{melon}, \mathit{grapes}, \mathit{strawberry} \}, 80,\\
&\hspace*{2.8cm}      \mathit{if}(s_1\approx s_2 \wedge s_2\approx s_3 \wedge s_1 \in \{ \mathit{bell}, \mathit{pear}, \mathit{blueberry}, \mathit{orange} \}, 40,\\
&\hspace*{2.8cm}      \mathit{if}(s_1\approx\mathit{star} \wedge s_2\approx \mathit{orange} \wedge s_3\approx\mathit{orange}, 40,\\
&\hspace*{2.8cm}      \mathit{if}(s_1\approx \mathit{pear} \wedge s_2\approx\mathit{star} \wedge s_3\approx \mathit{pear}, 40,\\
&\hspace*{2.8cm}      \mathit{if}(s_1\approx \mathit{blueberry} \wedge s_2\approx \mathit{blueberry} \approx s_3\approx \mathit{star}, 40,
                    0))))));\\
&     \mathit{reward}(i_1,i_2,i_3) = \mathit{price}(\mathit{get}_1(i_1{+}1), \mathit{get}_2(i_2{+}1), \mathit{get}_3(i_3{+}1)) +{}   &\textrm{\% line 1}\\
&\hspace*{2.95cm}       \mathit{price}(\mathit{get}_1(i_1{+}2), \mathit{get}_2(i_2{+}2), \mathit{get}_3(i_3{+}2)) +{}&\textrm{\% line 2}\\
&\hspace*{2.95cm}      \mathit{price}(\mathit{get}_1(i_1), \mathit{get}_2(i_2), \mathit{get}_3(i_3)) + {}&\textrm{\% line 3}\\
&\hspace*{2.95cm}        \mathit{price}(\mathit{get}_1(i_1), \mathit{get}_2(i_2{+}1), \mathit{get}_3(i_3{+}2)) +{}      &\textrm{\% line 4}\\
&\hspace*{2.95cm}       \mathit{price}(\mathit{get}_1(i_1{+}2), \mathit{get}_2(i_2{+}1), \mathit{get}_3(i_3));       &\textrm{\% line 5}\\
\textbf{act}  &\mathit{display}: \Nat;\\
\textbf{init} &
        \textbf{dist }i_1,i_2,i_3: \Nat[\mathit{distribution}(i_1)*\mathit{distribution}(i_2)*\mathit{distribution}(i_3)].\\
&        \hspace{4cm}\mathit{display}(reward(i_1,i_2,i_3)){\cdot}\delta;\\
\end{array}\]
\caption{The mCRL2 specification of the five-play-lines game}
\label{fiveplaylines}
\end{figure}

The behaviour of the five-play-line game is specified in mCRL2 in Figure \ref{fiveplaylines}. It describes
only a single run of this game, as only this is interesting. At the \textbf{init} it is stated that variables $i_1$, $i_2$ and
$i_3$ are chosen according to the distribution given above. Each $i_j$ is an index in reel $j$ smaller than 24, 
chosen with probability $\frac{1}{24}$. Based on the values $i_1$, $i_2$, $i_3$ a reward is displayed, which is 
defined using the mappings
$\mathit{reward}$ and $\mathit{price}$. The mapping $\mathit{reward}$ encodes the winning lines and the mapping
$\mathit{price}$ represents the price each winning line provides as indicated at the top of the slot machine. 

We are interested in the expected return to player of this game, given by the following formula, which only looks
at the payout of a single run. Recall that playing this game costs five credits. 
\[\sup n:\Nat.\langle \mathit{display}(n)\rangle (n-5).\]
This formula evaluates to $-0.2980$. As a game costs 5 credits, the loss per credit is $\frac{-0.2980}{5}=0.0596$.
Our analysis therefore yields a return to player of $94\%$, which is close to the RTP of 95\% reported on the website~\cite{OnlineGame}.

\subsection{The ten-play-lines game}
\begin{figure}
\[\begin{array}{ll@{\hspace{-1.6cm}}r}
\textbf{map }
&     \mathit{price}: \Nat\times\mathit{Symbol}\times\mathit{Symbol}|\times\mathit{Symbol}\rightarrow\Nat;\\
&     \mathit{reward}: \Nat\times\Nat\times\Nat\times\Nat \rightarrow \Nat;\\
\textbf{var}&  n, i_1, i_2, i_3, i_4: \Nat;\\
&     s_1, s_2, s_3: \mathit{Symbol;}\\
\textbf{eqn}&  \mathit{price}(n, s_1,s_2,s_3)=\mathit{if}(s_1\approx s_2 \wedge s_2\approx s_3,\\
&\hspace*{3.6cm}   \mathit{if}(s_1\approx\mathit{star}, 200,\\
&\hspace*{3.6cm}   \mathit{if}(s_1 \in \{ \mathit{seven}, \mathit{melon}, \mathit{grapes}, \mathit{strawberry} \}, 80,\\
&\hspace*{3.6cm}   \mathit{if}(s_1 \in \{ \mathit{bell}, \mathit{pear}, \mathit{blueberry}, \mathit{orange} \}, 40, 0))),\\
&\hspace*{3.2cm}\mathit{if}(n\approx 1,\\
&\hspace*{3.6cm}\mathit{if}(s_1\approx \mathit{star}\wedge s_2\approx \mathit{orange} \wedge s_3\approx \mathit{orange}, 40,\\
&\hspace*{3.6cm}  \mathit{if}(s_1\approx \mathit{pear} \wedge s_2\approx \mathit{star} \wedge s_3\wedge\mathit{pear}, 40,\\
&\hspace*{3.6cm}  \mathit{if}(s_1\approx \mathit{blueberry} \wedge s_2\approx\mathit{blueberry} \wedge s_3\approx \mathit{star}, 40, 0))),\\
&\hspace*{3.2cm}\mathit{if}(s_1\approx \mathit{star} \wedge s_2\approx \mathit{pear} \wedge s_3\wedge\mathit{pear}, 40,\\
&\hspace*{3.2cm}\mathit{if}(s_1\approx \mathit{blueberry} \wedge s_2\approx \mathit{star} \wedge s_3\approx \mathit{blueberry}, 40,\\
&\hspace*{3.2cm}\mathit{if}(s_1\approx \mathit{bell} \wedge s_2\approx \mathit{bell} \approx s_3\approx \mathit{star}, 40, 0)))));\\
&     \mathit{reward}(i_1,i_2,i_3,i_4) = \mathit{price}(1, \mathit{get}_1(i_1{+}1), \mathit{get}_2(i_2{+}1), \mathit{get}_3(i_3{+}1)) +{}        &\textrm{\% line  1~~}\\
&\hspace*{1cm}      \mathit{price}(1, \mathit{get}_1(i_1{+}2), \mathit{get}_2(i_2{+}2), \mathit{get}_3(i_3{+}2)) +{}        &\textrm{\% line  2~~}\\
&\hspace*{1cm}      \mathit{price}(1, \mathit{get}_1(i_1), \mathit{get}_2(i_2), \mathit{get}_3(i_3)) +{}              &\textrm{\% line  3~~}\\
&\hspace*{1cm}      \mathit{price}(1, \mathit{get}_1(i_1), \mathit{get}_2(i_2{+}1), \mathit{get}_3(i_3{+}2)) +{}          &\textrm{\% line  4~~}\\
&\hspace*{1cm}      \mathit{price}(1, \mathit{get}_1(i_1{+}2), \mathit{get}_2(i_2{+}1), \mathit{get}_3(i_3)) +{}          &\textrm{\% line  5~~}\\
&\hspace*{1cm}      \mathit{price}(2, \mathit{get}_2(i_2{+}1), \mathit{get}_3(i_3{+}1), \mathit{get}_4(i_4{+}1)) +{}        &\textrm{\% line  6~~}\\
&\hspace*{1cm}      \mathit{price}(2, \mathit{get}_2(i_2{+}2), \mathit{get}_3(i_3{+}2), \mathit{get}_4(i_4{+}2)) +{}        &\textrm{\% line  7~~}\\
&\hspace*{1cm}      \mathit{price}(2, \mathit{get}_2(i_2), \mathit{get}_3(i_3), \mathit{get}_4(i_4)) +{}         &\textrm{\% line  8~~}\\
&\hspace*{1cm}      \mathit{price}(2, \mathit{get}_2(i_2), \mathit{get}_3(i_3{+}1), \mathit{get}_4(i_4+2)) +{}          &\textrm{\% line  9~~}\\
&\hspace*{1cm}      \mathit{price}(2, \mathit{get}_2(i_2{+}2), \mathit{get}_3(i_3{+}1), \mathit{get}_4(i_4));           &\textrm{\% line 10}\\
\textbf{act}&  \mathit{display}: \Nat;\\
\textbf{init}&
    \textbf{dist }i_1,i_2,i_3,i_4: \Nat[\mathit{distribution}(i_1)*\mathit{distribution}(i_2)*\mathit{distribution}(i_3)*\mathit{distribution}(i_4)].\\
&\hspace*{3cm}    \mathit{display}(\mathit{reward}(i_1,i_2,i_3,i_4)).\delta;
\end{array}\]
\caption{The mCRL2 behaviour of the ten-play-lines game}
\label{tenplaylines}
\end{figure}
The ten-play-line game is very comparable to the five-play-line game. The differences are that playing it costs 10 credits,
but in return the payout is now in accordance to the 10 play lines indicated at the top of the cabinet. The additional five
play lines use columns B--D, and the overall payout is the sum of all winning sequences of three symbols over all 10 play lines.

The behaviour of this game is specified in Figure \ref{tenplaylines}. We use a parameter $n$ in $\mathit{price}(n,s_1,s_2,s_3)$
to indicate in which column the sequence $s_1$, $s_2$, $s_3$ starts. If $s_1$ is located in column 1 the winning combinations with
a $\mathit{star}$ are different than when $s_1$ is located in column 2. 

The formula for the expected average gain of this formula is the same as for the five-play-line game, except that each game now
costs 10 credits;
\[\sup n:\Nat.\langle \mathit{display}(n)\rangle (n-10).\]
Evaluating this formula yields $-0.5960$ which is twice the average expected loss for the five-play-lines game. This also 
yields a 94\% return to player, exactly the same as for the five-play-lines game. 
Apparently, there are no extra winning combinations for the 
ten-play-lines game from which a player can benefit.
This is in contrast to \cite{OnlineGame} which states a RTP of 97.50\% for the ten-play-lines game.

\subsection{The reel game}
\label{sec:reelgame}
\begin{figure}
\[\begin{array}{@{\hspace{-0.5cm}}ll@{\hspace{-1.6cm}}r}
\textbf{sort }&\mathit{Symbol} = \textbf{struct }\mathit{orange}\mid\mathit{grapes}\mid \mathit{pear} \mid\mathit{melon} \mid \mathit{blueberry} \mid \mathit{strawberry} \mid \mathit{bell} \mid\mathit{seven} \mid\mathit{star};\\
\textbf{map }&\mathit{check}: \mathit{Symbol}\times \mathit{Symbol}\times \mathit{Symbol}\times \mathit{Symbol} \rightarrow \Nat;\\
&    \mathit{checkwin}_4, \mathit{checkwin}_3, \mathit{checkwin}_2: \mathit{Symbol}\times \mathit{Symbol}\times\mathit{Symbol}\times\mathit{Symbol} \rightarrow \Bool;\\
\textbf{var }&s_1, s_2, s_3, s_4: \mathit{Symbol};\\
\textbf{eqn }&\mathit{check}(s_1, s_2, s_3, s_4) =\\
&\hspace*{1cm}\mathit{if}(\mathit{checkwin}_4(s_1,s_2,s_3,s_4), \\
&\hspace{2cm}\mathit{if}((s_1\approx \mathit{star} \wedge s_2\approx \mathit{star} \wedge s_3\approx \mathit{star} \wedge s_4\approx \mathit{star}), 200,\\
&\hspace*{2cm}\mathit{if}((s_1\approx \mathit{grapes} \vee  s_1\approx \mathit{melon} \vee  s_1\approx \mathit{strawberry} \vee  s_1\approx \mathit{seven}), 64, 20)),\\
&\hspace*{1cm}\mathit{if}(\mathit{checkwin}_3(s_1,s_2,s_3,s_4),\\
&\hspace*{2cm} \mathit{if}((s_1\approx \mathit{star} \wedge s_2\approx \mathit{star} \wedge s_3\approx \mathit{star}), 100,\\
&\hspace*{2cm}\mathit{if}((s_1\approx \mathit{grapes} \vee  s_1\approx \mathit{melon} \vee  s_1\approx \mathit{strawberry} \vee  s_1\approx \mathit{seven}), 16, 8)),\\
&\hspace*{1cm}\mathit{if}(\mathit{checkwin}_2(s_1,s_2,s_3,s_4),\\
&\hspace*{2cm} \mathit{if}((s_1\approx \mathit{star} \wedge s_2\approx \mathit{star}), 8,\\
&\hspace*{2cm}\mathit{if}((s_1\approx \mathit{grapes} \vee  s_1\approx \mathit{melon} \vee  s_1\approx \mathit{strawberry} \vee  s_1\approx \mathit{seven}), 4, 2)), 0)));\\
&    \mathit{checkin}_4(s_1, s_2, s_3, s_4) =\\
& \hspace*{2cm}(s_1\approx s_2 \wedge s_2\approx s_3 \wedge s_3\approx s_4) \vee {}\\
& \hspace*{2cm}(s_1\approx \mathit{star} \wedge s_2\approx \mathit{orange} \wedge s_3\approx \mathit{orange} \wedge s_4\approx \mathit{orange}) \vee{} \\
& \hspace*{2cm}(s_1\approx \mathit{pear} \wedge s_2\approx \mathit{star} \wedge s_3\approx \mathit{pear} \wedge s_4\approx \mathit{pear}) \vee{} \\
& \hspace*{2cm}(s_1\approx \mathit{blueberry} \wedge s_2\approx \mathit{blueberry} \wedge s_3\approx \mathit{star} \wedge s_4\approx \mathit{blueberry}) \vee{} \\
& \hspace*{2cm}(s_1\approx \mathit{bell} \wedge s_2\approx \mathit{bell} \wedge s_3\approx \mathit{bell} \wedge s_4\approx \mathit{star});\\
&    \mathit{check win}_3(s_1, s_2, s_3, s_4) =\\
& \hspace*{2cm}(s_1\approx s_2 \wedge s_2\approx s_3) \vee {}\\
& \hspace*{2cm}(s_1\approx \mathit{star} \wedge s_2\approx \mathit{orange} \wedge s_3\approx \mathit{orange}) \vee {}\\
& \hspace*{2cm}(s_1\approx \mathit{pear} \wedge s_2\approx \mathit{star} \wedge s_3\approx \mathit{pear}) \vee {}\\
& \hspace*{2cm}(s_1\approx \mathit{blueberry} \wedge s_2\approx \mathit{blueberry} \wedge s_3\approx \mathit{star});\\
& \mathit{checkwin}_2(s_1, s_2, s_3, s_4) =\\
& \hspace*{2cm}(s_1\approx s_2) \vee {}\\
& \hspace*{2cm}(s_1\approx \mathit{star} \wedge s_2\approx \mathit{orange}) \vee {}\\
& \hspace*{2cm}(s_1\approx \mathit{pear} \wedge s_2\approx \mathit{star});\\
\end{array}\]
\caption{The mCRL2 data description of the reel game}
\label{reelsgame1}
\end{figure}
\begin{figure}
\[\begin{array}{ll@{\hspace{-1.6cm}}r}

\textbf{act}&\mathit{play}: \Bool\times \Bool\times\Bool\times\Bool;\\
&     \mathit{points}:\Nat;\\
\textbf{glob }&\mathit{dc}:\mathit{Symbol};\\
\textbf{proc}&\mathit{Play}(\mathit{hold}_1, \mathit{hold}_2, \mathit{hold}_3, \mathit{hold}_4:\Bool, i_1, i_2, i_3, i_4: \mathit{Symbol}) =\\
&\hspace*{0.5cm}\mathit{play}(\mathit{hold}_1, \mathit{hold}_2, \mathit{hold}_3, \mathit{hold}_4){\cdot}\\
&\hspace*{0.5cm}\textbf{dist }s_1{:}\mathit{Symbol}[\mathit{if}(\mathit{hold}_1, \mathit{if}(i_1\approx s_1, 1, 0), \mathit{if}(s_1\approx \mathit{star},\frac{1}{24}, \mathit{if}(s_1\approx \mathit{orange},\frac{3}{24},\\
&\hspace{4.15cm}\mathit{if}(s_1\approx \mathit{grapes},\frac{2}{24},\mathit{if}(s_1\approx \mathit{pear},\frac{4}{24},\mathit{if}(s_1\approx \mathit{melon},\frac{2}{24},\\
&\hspace{4.15cm}\mathit{if}(s_1\approx \mathit{blueberry},\frac{4}{24},\mathit{if}(s_1\approx \mathit{strawberry},\frac{2}{24},\\
&\hspace{4.15cm}\mathit{if}(s_1\approx \mathit{bell},\frac{4}{24},\mathit{if}(s_1\approx \mathit{seven},\frac{2}{24}, 0))))))))))].\\
%%%%%%%%%%%%%%%
&\hspace*{0.5cm}\textbf{dist }s_2{:}\mathit{Symbol}[\mathit{if}(\mathit{hold}_2, \mathit{if}(i_2\approx s_2, 1, 0), \mathit{if}(s_2\approx \mathit{star},\frac{1}{24}, \mathit{if}(s_2\approx \mathit{orange},\frac{4}{24},\\
&\hspace{4.15cm}\mathit{if}(s_2\approx \mathit{grapes},\frac{2}{24},\mathit{if}(s_2\approx \mathit{pear},\frac{3}{24},\mathit{if}(s_2\approx \mathit{melon},\frac{2}{24},\\
&\hspace{4.15cm}\mathit{if}(s_2\approx \mathit{blueberry},\frac{4}{24},\mathit{if}(s_2\approx \mathit{strawberry},\frac{2}{24},\\
&\hspace{4.15cm}\mathit{if}(s_2\approx \mathit{bell},\frac{4}{24},\mathit{if}(s_2\approx \mathit{seven},\frac{2}{24}, 0))))))))))].\\
%%%%%%%%%%%%%%%%%
&\hspace*{0.5cm}\textbf{dist }s_3{:}\mathit{Symbol}[\mathit{if}(\mathit{hold}_3, \mathit{if}(i_3\approx s_3, 1, 0), \mathit{if}(s_3\approx \mathit{star},\frac{1}{24}, \mathit{if}(s_3\approx \mathit{orange},\frac{4}{24},\\
&\hspace{4.15cm}\mathit{if}(s_3\approx \mathit{grapes},\frac{2}{24},\mathit{if}(s_3\approx \mathit{pear},\frac{4}{24},\mathit{if}(s_3\approx \mathit{melon},\frac{2}{24},\\
&\hspace{4.15cm}\mathit{if}(s_3\approx \mathit{blueberry},\frac{3}{24},\mathit{if}(s_3\approx \mathit{strawberry},\frac{2}{24},\\
&\hspace{4.15cm}\mathit{if}(s_3\approx \mathit{bell},\frac{4}{24},\mathit{if}(s_3\approx \mathit{seven},\frac{2}{24}, 0))))))))))].\\
%%%%%%%%%%%%%%%%%%
&\hspace*{0.5cm}\textbf{dist }s_4{:}\mathit{Symbol}[\mathit{if}(\mathit{hold}_4, \mathit{if}(i_4\approx s_4, 1, 0), \mathit{if}(s_4\approx \mathit{star},\frac{1}{24}, \mathit{if}(s_4\approx \mathit{orange},\frac{4}{24},\\
&\hspace{4.15cm}\mathit{if}(s_4\approx \mathit{grapes},\frac{2}{24},\mathit{if}(s_4\approx \mathit{pear},\frac{4}{24},\mathit{if}(s_4\approx \mathit{melon},\frac{2}{24},\\
&\hspace{4.15cm}\mathit{if}(s_4\approx \mathit{blueberry},\frac{4}{24},\mathit{if}(s_4\approx \mathit{strawberry},\frac{2}{24},\\
&\hspace{4.15cm}\mathit{if}(s_4\approx \mathit{bell},\frac{3}{24},\mathit{if}(s_4\approx \mathit{seven},\frac{2}{24}, 0))))))))))].\\
%%%%%%%%%%%%%%%%%%
& \hspace*{0.5cm}( (\mathit{checkwin}_4(s_1,s_2,s_3,s_4) \vee  \mathit{checkwin}_3(s_1,s_2,s_3,s_4) \vee  \mathit{checkwin}_2(s_1,s_2,s_3,s_4))\\
& \hspace*{1cm}      \rightarrow
(\mathit{points}(\mathit{check}(s_1,s_2,s_3,s_4)){\cdot}\\
&\hspace*{1.55cm}           \mathit{Play}(\false, \false, \false, \false, \mathit{dc}, \mathit{dc}, \mathit{dc}, \mathit{dc}))\\
& \hspace{1cm}\diamond
~~~(\mathit{points}(\mathit{chec}k(s_1,s_2,s_3,s_4)){\cdot}\\
&\hspace*{1.55cm}\sum_{b_1, b_2, b_3, b_4: \Bool}.\mathit{Play}(b_1, b_2, b_3, b_4, s_1, s_2, s_3, s_4)));\\
\textbf{init }&\mathit{Play}(\false, \false, \false, \false, \mathit{dc}, \mathit{dc}, \mathit{dc}, \mathit{dc});
\end{array}\]
\caption{The mCRL2 process part of the reel game}
\label{reelsgame2}
\end{figure}
The reel game is played on four columns and a single line, where the hold buttons can be used. 
Playing this game costs one credit. 
The winning combinations are listed at the lower part of the cabinet. Essentially, a reward is awarded if the first 2, 3 or 4 symbols
from left to right are the same. So, for instance, four stars yield 200, three stars at the left yield 100 and two left stars provide
8 credits. As with the play-lines games, stars have a double role, namely as orange in the first column, as pear in the second, 
as blueberry in the third and as bell in the last columns. 

Interestingly, we had difficulties understanding how the hold buttons were supposed to be used. 
We think that it is useful to explicitly go through some variants we encountered, as it underlines the need for
a formal description of the game when communicating the rules to for instance gambling authorities. 
We summarize the expected payout for the different variants in Table~\ref{tab:reel_game}.

\begin{table}[b]
\centering
\caption{Comparison of variants for the reel game.}
\label{tab:reel_game}
\begin{tabular}{lrrr}
  \toprule
	Variant              & Max. expected gain & \multicolumn{2}{c}{RTP}\\
	                     & mCRL2 & mCRL2 & website\\
  \midrule
	Always allow hold       &  1.5652 & 256.52\% &     ---\\
	Hold with extra cost    &  0.6916 & 169.16\% &     ---\\ % with releasing hold buttons
	Only hold one round     & -0.1690 &  83.10\% & 88.50\%\\
  \bottomrule
\end{tabular}
\end{table}%

At a particular moment we got access to the playable version of the TopSpinner~\cite{OnlineGame},
which was helpful to increase understanding.
However, the average payouts mentioned on the website differ somewhat from our results. There are various explanations 
for this difference of which one is that we still do not exactly capture the rules of the game, 
although we consider this unlikely. 

\paragraph{Always allow hold buttons}
In our initial model, we assume that the hold buttons
can always be used, except at the start of the game and directly after a winning combination, as otherwise the 
winning combination could
be fixed, and a player could repeat this winning combination indefinitely. 

The data of this model is listed in Figure \ref{reelsgame1} and the process behaviour in Figure \ref{reelsgame2}. 
In the process behaviour
we use the $\textbf{glob}$ keyword to declare `global variables'. The meaning of these variables is that they can be set to any value,
without influencing the meaning of the specification. Tools instantiate these variables to appropriate values that allow to optimise
the verification efforts. 

The most interesting question that we like to ask is what the average maximal return to player is when the player plays the game optimally.
The following formula provides this maximal expected gain per game when the player plays $\mathit{max\_rounds}$ rounds:
\begin{equation}
\label{formulareelsgame}
\begin{array}{l}
\frac{1}{\mathit{max\_rounds}}*\mu X(\mathit{rounds}{:}\Nat{=}0).\\
\hspace*{2cm}            (\mathit{rounds}\approx \mathit{max\_rounds}) \wedge 0 \vee{}\\
\hspace*{2cm}              (\mathit{rounds}<\mathit{max\_rounds}) \wedge {}\\
\hspace*{3cm}\sup b_1,b_2,b_3,b_4{:}\Bool.\langle \mathit{play}(b_1,b_2,b_3,b_4)\rangle\\
\hspace*{4cm}               \sup n{:}\Nat.\langle \mathit{points}(n)\rangle (n{-}1+X(\mathit{rounds}{+}1))).
\end{array}\end{equation}

The variable $X(\mathit{rounds})$ in this formula has as value the maximal gain when $\mathit{rounds}$ games have been played. 
If the variable $\mathit{rounds}$ is equal to $\mathit{max\_rounds}$ the gain is set to $0$. Otherwise, a round of the game
takes place, modelled by the action $\mathit{play}(b_1,b_2,b_3,b_4)$ where the boolean $b_i$ is true when column 
$i$ is set to hold.
The formula calculates the supremum over all hold settings. It obtains the points $n$ that are won, and changes the
maximal gain by adding $n-1$ to the maximal gain. Note the use of $\sup n{:}\Nat.\ldots$. As the amount that is won is uniquely
defined, this supremum only acts as a binder for the variable $n$. 

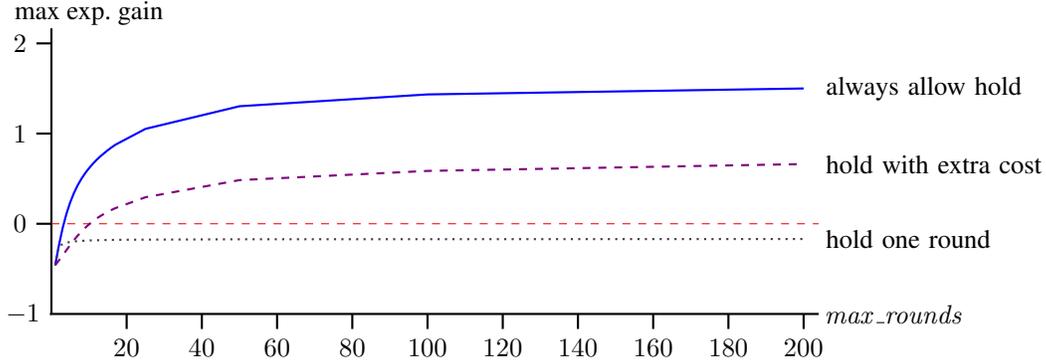
\begin{figure}
\begin{center}
\newcommand{\Y}[1]{1.2*#1+1.2}
\begin{tikzpicture}
%Axes
\draw [thick] (0,3.8) -- (0,0) -- (10.2,0);
\draw [thick] (1,0) -- (1,-0.2) node[below]{$20$};
\draw [thick] (2,0) -- (2,-0.2) node[below]{$40$};
\draw [thick] (3,0) -- (3,-0.2) node[below]{$60$};
\draw [thick] (4,0) -- (4,-0.2) node[below]{$80$};
\draw [thick] (5,0) -- (5,-0.2) node[below]{$100$};
\draw [thick] (6,0) -- (6,-0.2) node[below]{$120$};
\draw [thick] (7,0) -- (7,-0.2) node[below]{$140$};
\draw [thick] (8,0) -- (8,-0.2) node[below]{$160$};
\draw [thick] (9,0) -- (9,-0.2) node[below]{$180$};
\draw [thick] (10,0) -- (10,-0.2) node[below]{$200$};
\draw (11.2,0.0) node{$\mathit{max\_rounds}$};

\draw [thick] (0,\Y{-1}) -- (-0.02,\Y{-1}) node[left]{$-1$};
\draw [thick] (0,\Y{0}) -- (-0.2,\Y{0}) node[left]{$0$};
\draw [thick] (0,\Y{1}) -- (-0.2,\Y{1}) node[left]{$1$};
\draw [thick] (0,\Y{2}) -- (-0.2,\Y{2}) node[left]{$2$};
\draw [dashed,red] (0,\Y{0}) -- (10.2,\Y{0});
\draw (0.5,4.0) node{max exp.\ gain};
%Graph. x-axis is obtained by dividing by 20
%Graph of expected max gain after max_rounds  
\draw [thick, blue] (0.05, \Y{-0.46}) -- (0.1,-\Y{0.237}) -- (0.15, \Y{-0.045}) -- (0.2, \Y{0.111}) -- (0.25, \Y{0.238}) -- (0.3,\Y{0.342}) -- 
             (0.35,\Y{0.428}) -- (0.4,\Y{0.500}) -- (0.45,\Y{0.562}) -- (0.5,\Y{0.616}) -- (0.55,\Y{0.664}) -- (0.6, \Y{0.708}) -- 
             (0.65,\Y{0.747}) -- (0.7,\Y{0.782}) -- (0.75,\Y{0.816}) -- (0.8,\Y{0.847}) -- (0.85,\Y{0.876}) -- (1.25, \Y{1.052}) -- 
             (2.5,\Y{1.303}) -- (5,\Y{1.434}) -- (10,\Y{1.500});
\draw (11.8, \Y{1.500}) node[text width=3cm]{always allow hold};
             
\draw [thick, violet, dashed]
(0.05, \Y{-0.46}) -- (0.1,-\Y{0.406}) -- (0.15, \Y{-0.349}) -- (0.2, \Y{-0.292}) -- (0.25, \Y{-0.235}) -- (0.3,\Y{-0.181}) -- 
             (0.35,\Y{-0.131}) -- (0.4,\Y{-0.085}) -- (0.45,\Y{-0.044}) -- (0.5,\Y{-0.006}) -- (0.55,\Y{0.027}) -- (0.6, \Y{0.058}) -- 
             (0.65,\Y{0.085}) -- (0.7,\Y{0.110}) -- (0.75,\Y{0.133}) -- (0.8,\Y{0.154}) -- (0.85,\Y{0.173}) -- (1.25, \Y{0.294}) -- 
             (2.5,\Y{0.484}) -- (5,\Y{0.586}) -- (10,\Y{0.662});
\draw (11.8, \Y{0.662}) node[text width=3cm]{hold with extra cost};

\draw [thick, darkgray, dotted]
(0.05, \Y{-0.46}) -- (0.1,\Y{-0.237}) -- (0.15, \Y{-0.229}) -- (0.2, \Y{-0.209}) -- (0.25, \Y{-0.202}) -- (0.3,\Y{-0.196}) -- 
             (0.35,\Y{-0.193}) -- (0.4,\Y{-0.190}) -- (0.45,\Y{-0.187}) -- (0.5,\Y{-0.185}) -- (0.55,\Y{-0.184}) -- (0.6, \Y{-0.183}) -- 
             (0.65,\Y{-0.182}) -- (0.7,\Y{-0.181}) -- (0.75,\Y{-0.180}) -- (1.0,\Y{-0.177}) -- (1.25, \Y{-0.176}) -- 
             (2.5,\Y{-0.172}) -- (10,\Y{-0.169});             
	\draw (11.8, \Y{-0.169}) node[text width=3cm]{hold one round};
               
\end{tikzpicture}
\end{center}
\caption{The maximal expected gain per game for the reel game during $\mathit{max\_rounds}$ games}
\label{expectedgain}
\end{figure}

The results are unexpected and are provided in the graph in Figure \ref{expectedgain} as the continuous blue line. 
On the $x$-axis the
number of iteratively played games are indicated and on the $y$-axis the maximal average gain per game is depicted. When playing one
game the average loss is $-0.46$ credits, which is understandable as the player cannot influence the game with the hold buttons.
When playing at most two games the minimal loss is $-0.24$ per game. But when optimally playing four games there is already an average profit 
of $0.11$ credits. This increases to an average profit for the perfect player per game of more than $1.5$ credits if the game is played for more than 200 rounds. 

\paragraph{Hold with extra cost}
The question arises whether it is really possible to have an expected profit on this actual machine. 
After some discussion, the company suggested that when applying the hold buttons, an extra credit
had to be paid. This is easily modelled by changing the last part of Formula (\ref{formulareelsgame}) 
$n-1+X(\mathit{rounds}{+}1)$ into $n-\mathit{if}(b_1{\vee}b_2{\vee}b_3{\vee} b_4,2,1)+X(\mathit{rounds}{+}1)$.
The results are depicted in Figure \ref{expectedgain} as the middle purple dashed line. Clearly, the average gain is less,
but in the long run the player can still obtain a gain of $0.69$ per game. It is unlikely that this is actually the case. 

\paragraph{Only hold for one round}
After more discussion, and especially, looking at the slot machine simulator \cite{OnlineGame}, 
it appears that when the
hold buttons are used in a certain round, they cannot be used in the next round.

We model this by replacing the following line in the model in Figure~\ref{reelsgame2}:
\[\begin{array}{l}
\sum_{b_1, b_2, b_3, b_4: \Bool}.\mathit{Play}(b_1, b_2, b_3, b_4, s_1, s_2, s_3, s_4)
\end{array}\]
by
\[\begin{array}{l}
(\mathit{hold}_1{\vee}\mathit{hold}_2{\vee}\mathit{hold}_3{\vee}\mathit{hold}_4)\\
\hspace{2cm}                    \rightarrow \mathit{Play}(\false, \false, \false, \false, \mathit{dc}, \mathit{dc}, \mathit{dc}, \mathit{dc})\\
\hspace{2cm}                    ~\diamond\, \sum_{b_1, b_2, b_3, b_4: \Bool}\mathit{Play}(b_1, b_2, b_3, b_4, s_1, s_2, s_3, s_4).
\end{array}                    
\]
This expresses that if one of the hold button is pressed, the next round is played with all variables representing the hold buttons set to $\false$. Otherwise, if no hold button is pressed, the hold buttons in the next round can be set at will.

In the web simulator there is no additional fee for using the hold buttons, so we use Formula (\ref{formulareelsgame}) without alteration.
Evaluating Formula (\ref{formulareelsgame}) yields rewards conforming to the lower, grey dotted line in Figure \ref{expectedgain}. 
On the long run, when playing optimally, a loss of $-0.17$ per game is expected, i.e., a return to player of $83\%$.
This result is not in line with the stated RTP of $88.5\%$ on the website~\cite{OnlineGame}.

\paragraph{Validation}
Comparing the RTP from our mCRL2 model with the RTPs stated on the website~\cite{OnlineGame}, we see discrepancies.
This is similar to the five-play-lines and ten-play-lines game where the website claims RTPs of respectively $95\%$ and $97.5\%$, whereas we find an RTP of $94\%$ for both games. 
In order to exclude mistakes from our side, we also modelled and analysed the games using the model checkers Storm~\cite{Storm} and Prism~\cite{Prism}.
Both tools yield the same results as mCRL2.

\paragraph{Optimal strategies}
\begin{figure}[t]
\centering
\includegraphics[height=12cm]{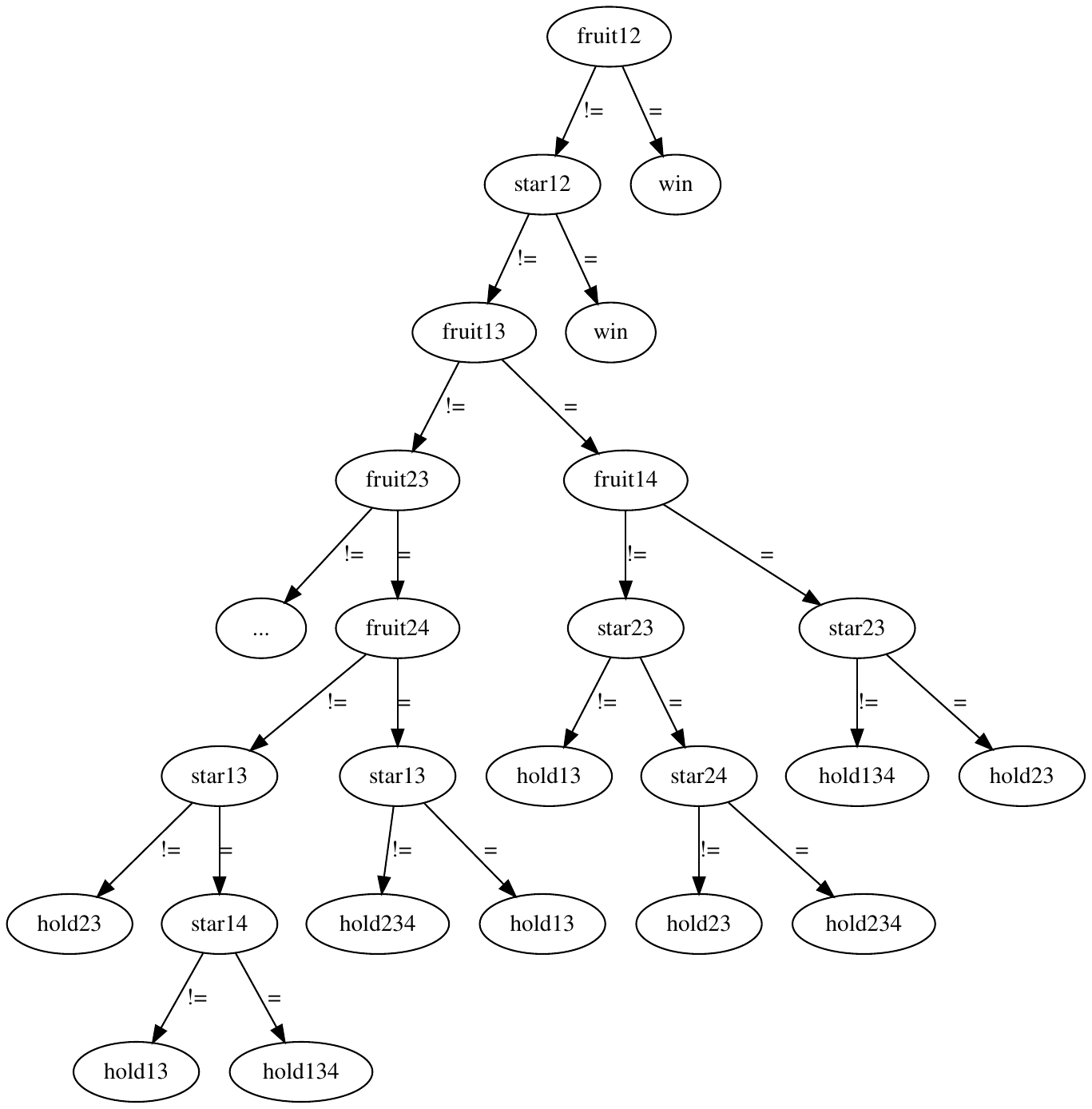}
\caption{Decision tree representing (parts of) the optimal player strategy for the reels game where a button holds only for one round.}
\label{fig:strategy}
\end{figure}
An interesting question as a player is when and which buttons to hold in order to achieve the best RTP.
For the reel game in which the buttons can only be held for one round, we can extract the player strategy leading to the optimal RTP of 83\%.
To this end, we use the Storm model checker~\cite{Storm} and export the optimal strategy which details for each possible reel outcome which buttons should be held.
In order to achieve a small and human understandable strategy, we employ the tool dtControl~\cite{DBLP:conf/tacas/AshokJKWWY21} to create a decision tree from the obtained strategy.

Figure~\ref{fig:strategy} depicts a part of the decision tree for the reels game in which a button can only be held for one round.
A node of the form `fruit$\mathit{x}\mathit{y}$' represents the check whether reels $x$ and $y$ show the same fruit symbol, `star$\mathit{x}\mathit{y}$' checks whether both reels show the star symbol.
The player strategy is given by `hold$\mathit{x}\mathit{y}\mathit{z}$' which represents that the player should hold the indicated columns.

The obtained strategy leads to some interesting insights.
For instance, stars are higher valued than fruits.
This can for instance be seen in the case where reels 1, 3 and 4 show the same fruit symbol (`fruit13', `fruit14') 
but reels 2 and 3 also show stars (`star23').
The outcome then is `hold23' meaning that the player should hold the buttons for the two stars instead of the three buttons for the equal fruits.
Similarly, if stars are present in reels 1 and 3, and the fruits in reel 2, 3, and 4 are equal, then the buttons 1 and 3 for the stars should be held.

\section{Conclusion}
\label{sec:conclusion}
We analysed the behaviour of slot machines and came to the conclusion that process specification and modal formulas are a suitable set of means to analyse them.
Once set up, a formal model of the slot machine has several benefits.
First, there is a clear and unambiguous description of the behaviour of the slot machine.
Second, the specifications and formulas can easily be adapted to investigate variations of the slot machines or particular scenarios that players could employ.
Third, the analysis via model checking computes exact values and therefore provides verified results.
Fourth, the approach is compositional and therefore allows to model a complete slot machine by combining the models for the individual games.
And finally, the player strategies can automatically be extracted, which allows better insights into the game mechanics.

While modeling the slot machines, we observed that it is not very easy to figure out what the exact rules of games are, not even for `simple' slot machines.
This is in line with what we found in board games as well \cite{DBLP:journals/siamrev/GrooteWZ16}.
It would therefore be useful to describe such games in a standardised formal way.
This allows to communicate the rules of the games with gaming authorities that 
verify that the return-to-player is within acceptable and legal bounds.

\subsection{Future work}
There are a quite a few interesting aspects of slot machines that deserve further investigation. 

Apart from the RTP, 
we could check whether the loss per hour is limited, even when a player employs a particularly bad strategy.
Another question could be what the game characteristics are, when the player has a limited amount of money or a limited
amount of rounds left. Note that the approach with quantitative modal logics is quite suitable to address such questions. 

Using various tools we derived explicit optimal strategies for players, especially for variations of the reel game.
These strategies are remarkably complex. In particular these strategies can take into account how many games the player is still planning to play. 
As yet, there are no easy ways to extract such strategies in the mCRL2 toolset, comparable to finding evidence
and counter examples as can be provided for non-quantitive model checking \cite{counterexample}. 
It would be very useful to extend the quantitive checkers of mCRL2 with such capabilities, similar to e.g.,~\cite{Storm,Prism}.

Another issue that can occur is that it can be computationally complex to evaluate a formula. This is essentially done
by translating the specification and the formula to a parameterised real equation system that is subsequently solved \cite{GrooteWillemse2023}
very much in line with how boolean formulas are solved using boolean equation systems \cite{GM14}. But the size of these equation systems
can grow rapidly. Solving formula (\ref{formulareelsgame}) for $\mathit{max\_rounds}=200$ requires more than 1M pres variables. 
Typical solutions here lie in using symbolic methods that can handle much larger systems, such as \cite{symbolic} for non-probabilistic and \cite{DBLP:journals/sttt/KwiatkowskaNP04,DBLP:phd/dnb/Hensel18} for probabilistic models.

\paragraph{Data availability.}
The process descriptions of slot machines and the formulas are provided in the examples directory of
the mCRL2 toolset (\url{www.mcrl2.org}) which is distributed under the liberal BOOST licence allowing its free use.

\bibliographystyle{plain} % We choose the "plain" reference style
\bibliography{references}
\end{document}